\documentclass[10pt]{article}

\usepackage{arxiv}

\usepackage[utf8]{inputenc} 
\usepackage[T1]{fontenc}    
\usepackage{hyperref}       
\usepackage{url}            
\usepackage{booktabs}       
\usepackage{amsfonts}       
\usepackage{nicefrac}       
\usepackage{microtype}      
\usepackage{lipsum}		
\usepackage{graphicx}
\usepackage{doi}
\usepackage[normalem]{ulem}
\usepackage[toc,page]{appendix}

\usepackage{amsmath}
\usepackage{amsthm}
\usepackage{amssymb}
\usepackage{stmaryrd}
\usepackage{soul}
\usepackage[
backend=biber,
style=numeric,
sorting=none
]{biblatex}

\theoremstyle{plain}
\newtheorem*{prop*}{\protect\propositionname}
\providecommand{\propositionname}{Proposition}

\title{Strong Coupling Expansions from a Fourier Duality in Scalar Lattice QFT}

\author{Nikita A. Ignatyuk$^*$\\
	Moscow Institute of Physics\\
	  and Technology (MIPT),\\
	Skolkovo Institute of \\
    Science and Technology (Skoltech)\\
	\texttt{ignatyuk.na@phystech.edu} \\
	\And 
    Daniel V. Skliannyi$^*$\\
	Weizmann Institute of Science\\
    \texttt{daniil.skliannyi@weizmann.ac.il} \\
    \thanks{These authors contributed equally to this work}
}

\renewcommand{\shorttitle}{Strong Coupling Expansions in Scalar Lattice QFT}

\hypersetup{
pdftitle={Strong Coupling Expansions from a Fourier Duality in Scalar Lattice QFT},
pdfsubject={mphys},
pdfauthor={Nikita A. Ignatyuk, Daniel Skliannyi},
pdfkeywords={Quantum field theory, Duality, Path integration, Lattice Field Theory},
}
\usepackage{xcolor}

\begin{document}
\maketitle

\begin{abstract}
Dualities between quantum field theories provide useful descriptions of
otherwise inaccessible parameter regimes. We develop a strong-coupling
expansion for a class of Euclidean scalar field theories on a lattice by
applying a Fourier transform to the local interaction
term. We focus on a self-interacting $\phi^4$ theory on a (periodic) hypercubic
lattice in arbitrary dimension and derive a dual representation in which
the strong-coupling regime of the original model is described by weak
interactions of a generally nonlocal dual field. Using standard
diagrammatic techniques, we obtain partially resummed approximations for
the free-energy density and the momentum-space two-point function,
including dual interaction vertices through nominal order $g^{-{10}}$.
For $d=2$ and $d=3$, the resulting expressions agree well with
Hamiltonian Monte Carlo simulations over the parameter ranges studied and
provide complementary approximations with an overlap in the weak-to-intermediate coupling region. We also discuss
the assumptions and limitations of the construction and illustrate its
application to the Ising model.
\end{abstract}
\newpage{}

\tableofcontents{}

\newpage{}

\section{Introduction}
\label{sect:intro}
Quantum Field Theory (QFT) \cite{Peskin-Shroeder,Bjorken:1965zz,Kleinert-phi-4, Mussardo:SQFT, Ryder_1996, Itzykson_Drouffe_1989, Itzykson:QFT,Wipf:SQFT} is a foundation of modern theoretical physics, providing a mathematical framework for the understanding of fundamental particles and their interactions. In this paper, we focus on the scalar Euclidean Quantum Field Theory on a lattice and develop some new techniques to obtain strong coupling expansions.

In this paper, we restrict our attention to the lattice field theories \cite{Smit_2002-lattice, morningstar2007monte-lattice, Wipf:SQFT, Modern_Perspectives_in_Lattice_QCD} because here the notion of path integral is mathematically well-defined (at least for the case of finite lattices). As a result, one can perform well-defined finite-dimensional calculations, avoiding the ambiguities of formal manipulations with continuum path integral. In the case of lattice field theory, the accuracy of the results can be independently benchmarked using numerical simulations. This is substantially more difficult in the case of continuum theories, since the computation formulas here are usually more sophisticated \cite{Lobanov1997, Hari_2023}. Another approach is to use some lattice simulations with the subsequent scaling of parameters and some kind of extrapolation in the lattice spacing parameter.

For simplicity, we fix the lattice spacing and won't discuss renormalization problems, expressing everything via the bare parameters of an action. The problem, nevertheless, is still physically interesting because it provides insights into the partition function structure and can be extended to the renormalization group approach. The renormalization-group flow of lattice theories is another important direction, but its analysis lies beyond the scope of the present work.

Given the restrictions above, the Lattice Euclidean QFT for the scalar field will be analyzed more closely. The most common technique for calculating correlation functions is Feynman Perturbation Theory, obtained by the expansion of an exponent with interaction in power series. This usually produces an expansion in positive powers of coupling constant, which is generally asymptotic rather than convergent. Typically, these series diverge rather fast and require the resummation techniques, e.g. Borel resummation. We will refer to this series as the Weak Coupling Perturbation Series in this paper.

In addition to the weak coupling perturbation series, strong coupling expansions and extensions of the weak series to the intermediate coupling regime have been widely studied. These approaches include:
\begin{enumerate}
    \item Resummation and analytic continuation techniques \cite{Kleinert-phi-4,Kleinert:1998-strong-coupling,Suslov_2001,Kazakov:1978ey,Kubyshin1984SommerfeldWatsonSO}. Their implementation generally requires high-order perturbative input and additional analytic information about the large-order behavior of the series;
    \item The closest earlier construction to the approach presented here is the kinetic-term expansion of ~\cite{strong_coupling_Bender}. Here, the interacting local measure is treated exactly, while the kinetic term is generated perturbatively by differential operators. This yields strong-coupling expansions at finite momentum cutoff and was also used to explore the approach to the continuum limit. The resulting perturbative organization differs from the field-space Fourier representation developed here, in particular in the treatment of the quadratic part and in the associated diagrammatic expansion. In addition, the individual terms of the resulting strong-coupling series are singular as $g\to0$, and the diagrammatic organization differs from the standard Feynman-diagram expansion used in the present work;
    \item Related strong-coupling constructions based on particular nonlinear classical solutions and saddle-point methods were developed in Refs.~\cite{FRASCA_2007_infrared,FRASCA_2007_phir_triv,FRASCA_2007_PT_duality}. The resulting expansion depends on a specific classical background, which must be chosen appropriately. Higher-order terms also become increasingly complicated analytically;
    \item Hopping-parameter expansions \cite{Montvay1997-be,Smit_2002-lattice} provide a natural lattice strong-coupling series in which the intersite couplings are treated perturbatively. They are therefore conceptually close to kinetic-term strong-coupling expansions, although their organization is different and sometimes involves complicated geometrical intuition in higher orders;
    \item The strong-coupling (or low-temperature) expansions based on Kramers--Wannier duality are used in lattice spin systems, such as the Ising model \cite{Kramers-Wannier-Poisson, Ising-low-t-decompositions-I-G-Enting-1994,Krammers-Wannier-duality,Lattice-Ising-JOitmaa,Onsager-Ising,Wipf:SQFT}. However, they rely on the specific discrete structure of the model and do not transfer directly to generic scalar lattice field theories. They also may require complicated combinatorial considerations for clusters calculation in higher orders;
    \item Padé approximants and related rational approximations \cite{Kleinert-phi-4,Bessis1968-sh} can extend perturbative information beyond the weak-coupling regime, although their accuracy depends on the available perturbative input and the chosen approximant. Besides, this approximant chose usually has no clear physical intuition behind.
\end{enumerate}

A related auxiliary-field formulation was recently considered in \cite{chen2026convergentkinetictermperturbationexpansion}. It is closely related to the field-space Fourier representation developed here, but differs in the perturbative organization and in the observables emphasized.

There are also a variety of other works that are beyond the scope of this paper. The approach presented in this paper proposes a way to connect the advantages of the aforementioned techniques. 

In this work, we derive an exact field-space Fourier representation of the generating functional for a broad class of scalar lattice field theories. In the strong-coupling regime, the resulting dual formulation admits an expansion in inverse powers of the original coupling constant. By incorporating the quadratic term of the dual interaction into the Gaussian part of the action, we obtain a diagrammatic strong-coupling expansion with a dressed dual propagator. For the quartic interaction, we derive explicit expansions for the free energy per lattice site and the momentum-space two-point function, and benchmark the resulting finite-order approximations against independent Hamiltonian Monte Carlo simulations in two and three dimensions.

The duality considered here is an exact transformation of the finite-lattice generating functional into a representation in terms of field-space Fourier-dual variables. We do not assume a local self-duality or an equivalence of continuum fixed points. 

The resulting dual formulation leads to finite-order approximations that are
useful in both the strong- and intermediate-coupling regimes studied below.
In contrast to the kinetic-term expansion of Ref.~\cite{strong_coupling_Bender},
the present formulation admits a standard Feynman-diagram organization, and
the individual finite-order terms remain regular as $g\to0$. This regularity
should not be interpreted as evidence of convergence or of accuracy in the
weak-coupling regime; rather, the numerical results below show that the
strong-coupling truncations remain quantitatively useful into the
intermediate-coupling region.

The paper is organized as follows. In Section~\ref{sect:duality-derivation}, we introduce the necessary definitions, derive the general field-space Fourier duality for scalar lattice theories, and specialize it to power-law interactions. In Section~\ref{sect:free-energy}, we derive the strong-coupling expansion of the free energy per lattice site, Eq.~\eqref{free-energy-per-site-def}, and compare it with the weak-coupling expansion and numerical simulations described in Appendix~\ref{appendix:num-methods}. In Section~\ref{sect:two-point-func}, we obtain the corresponding strong- and weak-coupling expansions of the momentum-space two-point function and benchmark them numerically. Section~\ref{sect:discussion} discusses the results and their limitations, and Section~\ref{sect:conclusions} summarizes the main conclusions and possible extensions.

\section{Duality derivation}
\label{sect:duality-derivation}
The main purpose of this paper is to present a convenient construction of the strong-coupling expansions in some Lattice QFTs. We do focus on the numerical correspondence of the obtained analytical expansions with simulation results rather than on exploring qualitatively new phenomena since the models we consider are fairly well-studied. We fix lattice spacing (setting it equal to $1$ in our system of units) and we do not treat any renormalization problems. The only limit we will consider is the thermodynamic limit.

\subsection{Lattice model and conventions}

Let us begin with the scalar theory on a cubic Euclidean $d$-dimensional lattice with periodic boundary conditions and $M$ sites in every dimension with a power-law potential and the action of the following form:

\begin{equation}
\label{lattice-action}
    S[\phi]=\frac{1}{2}\sum_{x, x'}L_{x,x'}\phi(x)\phi(x')+\sum_{x}V(\phi(x)),\qquad V(\phi)=\frac{1}{(2n)!}g^{2n}\phi^{2n},
\end{equation}
for some integer $n > 1$. Here all coordinate sums go over all lattice sites, whose total number we denote by $N=M^d$. In \eqref{lattice-action} $L_{x,x'}$ are the matrix elements of the operator:
\begin{equation}
    L = -\alpha \triangle + \gamma
\end{equation}
 where $\alpha$ and $\gamma$ are some constants greater than zero, and: 
\begin{equation}
\label{discrete-laplacian}
    (\triangle f)(\vec{r})=\sum_{j=1}^{d}(f(\vec{r}+\vec{e}_{j})-2f(\vec{r})+f(\vec{r}-\vec{e}_{j})),
\end{equation}
 is the lattice Laplacian. Vector $\vec{e}_{j}$ is the $j$th element of an orthonormal frame, corresponding to the cubic lattice with unit lattice spacing in each direction. This formula is a finite-difference discretization of the continuum Laplacian.

 It is the simplest possible model, which also serves as the building block for more complicated models. Consequently, we use it to test the core aspects of strong coupling and numerical approaches. This allows us to compare our results with other works and well-established numerical simulations to understand the strengths and limitations of the approach discussed in this manuscript. In addition, in Sections \ref{sect:free-energy}, \ref{sect:two-point-func} we will put $n=2$, which leads us to lattice $\phi^4$ theory.

For periodic boundary conditions on a cubic lattice with $M$ sites in each dimension, the eigenvalues $\{\lambda_{k} \}$ and eigenvectors $\{ h_k (x)\}$ of the operator $L$ have the form described in Appendix~\ref{appendix:laplacian}:

\begin{equation}
\label{laplacian-eigensystem}
\lambda_{k} = \gamma + 4\alpha \sum_{j=1}^{d} \sin^{2}\left(\frac{p_{j}}{2}\right),\qquad\vec{p}_{k}= \frac{2\pi\vec{k}}{M},\qquad\vec{k}\in \{0,\ldots,M-1\}^{d},\qquad h_{k}(x)=\frac{1}{\sqrt{N}}e^{i\left< p_{k},x \right>}.
\end{equation}

Here $p_k$ are conventionally called admissible lattice momenta by analogy with plane waves. Together with the quadratic operator $L$ it is useful to consider its Green's function $G$:

\begin{equation}
\label{Green-func-def}
    G = L^{-1},
\end{equation}
and in terms of the introduced eigenvalues and eigenfunctions these operators have the representations:
\begin{equation}
\label{green-func-raw}
G_{x,y}=\sum\limits_{k}\frac{h_{k}(x)h_{k}^*(y)}{\lambda_{k}}, \qquad L_{x,y}=\sum\limits_{k} \lambda_{k} h_{k}(x)h_{k}^*(y).
\end{equation}
One can also expand any lattice function $\phi(x)$ in terms of the operator $L$ eigenvectors, that is:
\begin{equation}
\label{dft-def-first}
\phi(x)=\sum\limits_{k}\phi_{k} h_{k}(x),\quad\phi_{k}=\sum\limits_{x} \phi(x) h_{k}^*(x).
\end{equation}

We fix the notation $\phi_k$ for the component of field $\phi(x)$ along the eigenvector $h_{k}(x)$. In the following part of the paper by the Discrete Fourier Transform we will mean similar relation but with a different normalization for greater convenience.

We will use the following notation for the Discrete Fourier Transform (DFT) throughout this paper. Given the lattice function $H(x)$, we will call its DFT the momentum-space function $H(p)$ satisfying the relation:

\begin{equation}
\label{DFT-inverse-finite}
    H(x) = \frac{1}{ N} \sum\limits_{p} H(p) e^{ipx}.
\end{equation}

Let us underline the notation difference between $H(p)$ and $H_p$. The inverse formula is straightforward:

\begin{equation}
\label{DFT}
    H(p) = \sum\limits_{x} H(x) e^{-ipx}.
\end{equation}
 
Now, for $N\gg 1$  the sum in \eqref{DFT-inverse-finite} becomes the integral:

\begin{equation}
\label{DFT-inverse}
    H(x) = \frac{1}{(2\pi)^d} \int\limits_{[0;2\pi]^d} H(p) e^{ipx} d^d p.
\end{equation}

One can see that in the chosen Fourier-transform convention, the discrete momentum sum approaches the standard Brillouin-zone integral in the thermodynamic limit. This relation will be helpful for the study of the correlation functions in momentum representation.

\subsection{Partition function and observables}
We define the partition function in an external source $j(x)$ as:

\begin{equation}
    \label{PartFunc1}
    \mathcal{Z}[j]=\int_{\mathbb{R}^{N}}\frac{\prod_{x}d\phi(x)}{\sqrt{(2\pi)^ N \det G}}\ \exp\left[-S[\phi]+ i\left\langle j,\phi\right\rangle \right],
\end{equation}

where $S[\phi]$ is the lattice action \eqref{lattice-action}. This partition function is inspired by Ising or XY-models \cite{Mussardo:SQFT}, being the integration over all system's states. In this paper we will focus mainly on the free energy density, which means $j=0$, but for generality we provide a form of duality for general fields $j$.

For the given partition function, we understand the correlation functions or, equivalently, correlators as:

\begin{equation}
\label{correlators-def-initial}
    \left \langle \phi(x_1) \cdot \ldots \cdot \phi(x_k) \right \rangle = \frac{1}{\mathcal{Z}[0]} \int_{\mathbb{R}^{N}}\frac{\prod_{x}d\phi(x)}{\sqrt{(2\pi)^ N \det G}} \ \phi(x_1) \cdot \ldots \cdot \phi(x_k) \ \exp \left(-S[\phi]\right),
\end{equation}

which is a common definition for Quantum Field Theory \cite{Mussardo:SQFT,Peskin-Shroeder}. Applying \eqref{DFT} to every $x_i$ dependence, one obtains $k$-point correlation function in the momentum representation, which is a definition of this object. Feynman Perturbation Theory for such correlators is well-known \cite{Ryder_1996,Peskin-Shroeder,Mussardo:SQFT,Itzykson:QFT,Kleinert-phi-4,Wipf:SQFT} and will be considered as a prerequisite in the following part of the paper.

The most important case for the present study is $k=2$, so let us explicitly write down the expression for the two-point correlation function in the momentum representation as Discrete Fourier Transform \eqref{DFT} of coordinate correlation function \eqref{correlators-def-initial}:

\begin{equation}
\label{two-point-momentum-def}
    \langle \phi(p) \phi(-p) \rangle = \sum\limits_x \langle \phi(0) \phi(x) \rangle e^{-ipx},
\end{equation}

where we have used the translational invariance of the theory \eqref{lattice-action}.

Note that we have chosen the special normalization such that $\mathcal{Z}=1$ for $j=0$ and $g=0$. It is not only a convenience but a necessary step to consider further the thermodynamic limit. Also, for the case of free theory, when $V(\phi)=0$, the partition function can be easily calculated:

\begin{equation}
\label{partition-func-free}
   \mathcal{Z}_0[j(x)] = \exp \left(-\frac{1}{2}\sum_{x,x^{'}}j (x) G_{x,x^{'}}j(x^{'})\right)
\end{equation}

Moreover, throughout the paper we will need the notion of a free energy, which we will identify up to a sign with a Generating Functional of Connected Green's Functions. Namely, we define free energy functional $\mathcal{F}[j(x)]$ as:

\begin{equation}
\label{free-energy-def}
    \mathcal{Z}[j(x)] = \exp\left(-\mathcal{F}[j(x)]\right).
\end{equation}   

Obviously, if one uses Feynman diagrams to calculate $\mathcal{Z}[j(x)]$, then the value of $\mathcal{F}[j(x)]$ will be given by the sum of connected diagrams \cite{Itzykson:QFT,Peskin-Shroeder,Ryder_1996,Wipf:SQFT,Mussardo:SQFT}.

We are mainly interested in Free energy density per site in thermodynamic limit with normalization with respect to free theory, which we define as:

\begin{equation}
\label{free-energy-per-site-def}
    f = \lim\limits_{N\rightarrow\infty} \frac{\mathcal{F}[0]}{N}.
\end{equation}

We will not prove analytically that this limit exists. For the purposes of this study, a numerical confirmation of finiteness suffices. As an additional verification, let us note that every term in both strong and weak perturbation theory we deduce later gives the finite contribution to $f$. Keeping in mind the link between the Ising model and Quantum Field Theory from a perspective described in Appendix~\ref{appendix:duality-for-Ising}, one can also draw the parallel with the Ising model free energy per site \cite{Baxter:1982zz}, also staying finite in thermodynamic limit.

The action considered is translationally invariant, so we denote:

\begin{equation}
\label{GF-shorthand-notations}
    G_{x,x}=G_{0}, \qquad  G_{x,0}=G_{x}
\end{equation}

Generally, due to the translational invariance of the type of propagators considered, one can write $G_{x+a,y+a}=G_{x,y}$. So we can introduce the propagator as a function of one variable as:

\begin{equation}
    G(x-y) = G_{x-y} = G_{x,y}.
\end{equation}

Now we consider the thermodynamic limit $N \gg 1$. 

We are able to write down a more convenient expression for the Green's function, assuming that the site number $N\gg 1$, as is commonly done in lattice field theories \cite{Montvay1997-be}:

\begin{equation}
\label{weak-propagator}
G_{x,x^{\prime}} =\left(L^{-1}\right)_{x,x^{\prime}}=\frac{1}{ \left(2\pi\right)^{d}}\left(\prod_{i=1}^{d}\int_{0}^{2\pi}dq_{i}\right)\left(4\alpha\sum_{j=1}^{d}\sin^{2}\left(\frac{q_{j}}{2}\right)+\gamma \right)^{-1} e^{-i\left(q,(x-x^{\prime})\right)}.
\end{equation}
This equality can be obtained from \eqref{green-func-raw}  by substituting the explicit lattice Laplacian eigenvalues and eigenvectors \eqref{laplacian-eigensystem} with a subsequent application of the Euler--Maclaurin summation formula. For convenience, we will also use the following notation for the Fourier Transform of $G_{x,x'}$ in the sense of \eqref{DFT-inverse}, which we will denote as $G(q)$:

\begin{equation}
\label{weak-propagator-ft}
    G(q) =  \left(4\alpha \sum_{j=1}^{d}\sin^{2}\left(\frac{q_{j}}{2}\right)+\gamma\right)^{-1}.
\end{equation}
The DFT $G(q)$ will be distinguished from $G(x-y)$ by the notation of its argument. That is, we will continue denoting the coordinates with Latin letters $x,\ y,\ z$, and we will use letters $p,\ q, \ k$ for the lattice momenta.

\subsection{Field-space Fourier duality}

There are several equivalent ways to derive the desired duality, but we will present the simplest one. 

For the interaction of one lattice site, let us use the Fourier transform for dimensionless interaction:

\begin{equation}
\label{HS-duality}
e^{- \frac{\phi^{2n}}{(2n)!}} = \frac{1}{2\pi}\int \limits_{\mathbb{R}} d \psi \, \tilde{V}(\psi) e^{i\phi \psi}, 
\end{equation}

where $\tilde{V}$ is a Fourier Transform of $e^{- \frac{\phi^{2n}}{(2n)!}}$:

\begin{equation}
\label{Interaction-Fourier-Transform}
\tilde{V}(\psi)=\int \limits_{\mathbb{R}} d\phi \, e^{- \frac{\phi^{2n}}{(2n)!} - i \phi \psi}.
\end{equation}

\begin{figure}
\begin{centering}
\includegraphics[width=8cm,height=5cm]{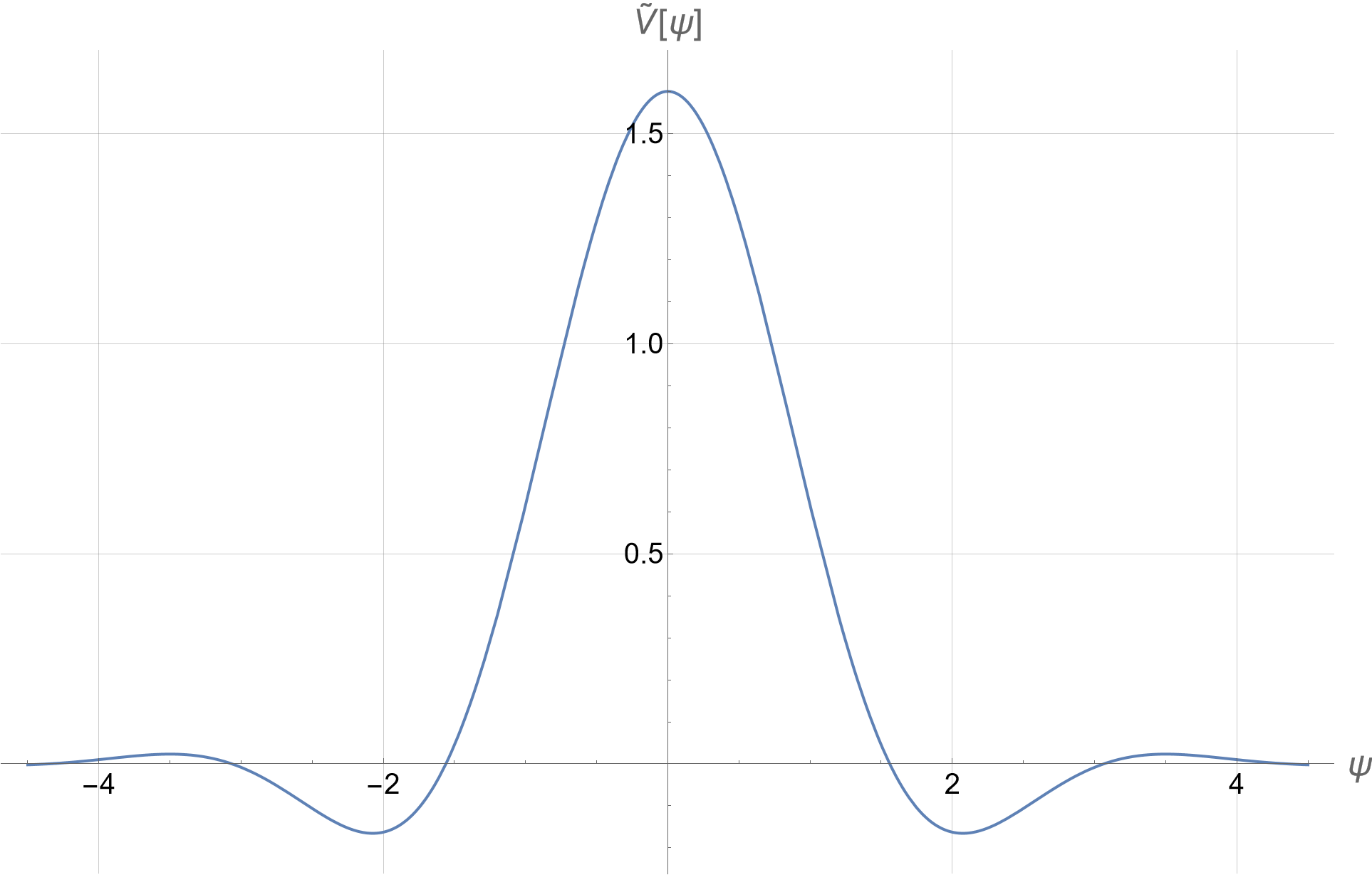}
\hfill
\includegraphics[width=8cm,height=5cm]{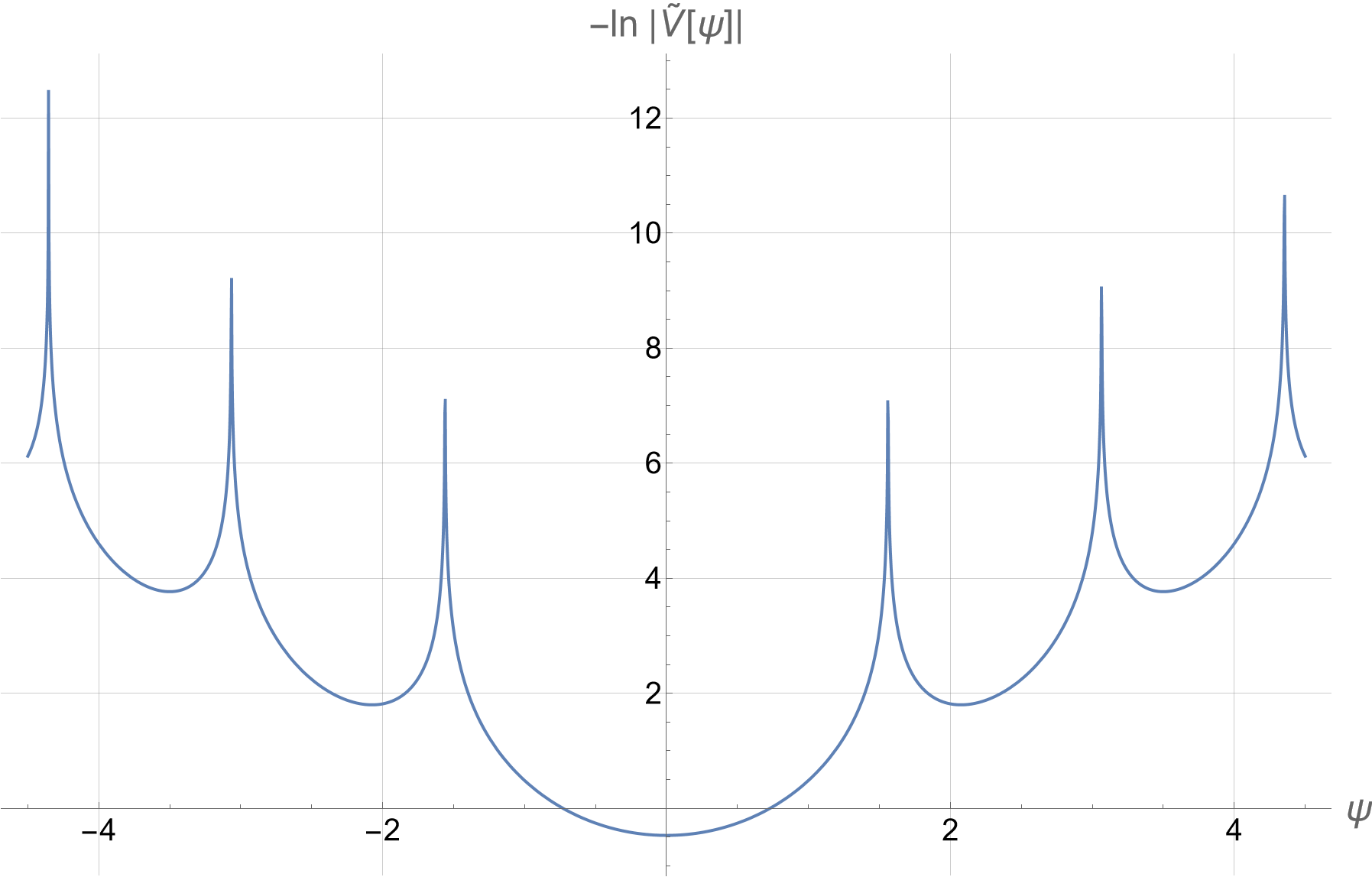}
\par\end{centering}
\caption{Plots of $\tilde{V}(\psi)$ \eqref{HS-duality} and $-\ln \left|\tilde{V}(\psi) \right|$.}
\label{graph:dual-potential}
\end{figure}

The plots of $\tilde{V}(\psi)$ and $-\ln \left|\tilde{V}(\psi) \right|$ are shown in Figure~\ref{graph:dual-potential}. In the following we will need a power series expansion for $\ln\tilde{V}$ of the form:

\begin{equation}
\label{dual-interaction-log-expansion}
    \ln \tilde{V}(\psi) = \ln \tilde{V}(0) - \frac{a}{2} \psi^{2}-\frac{b}{24}\psi^{4}-\frac{c}{6!} \psi^{6} -\frac{d}{8!} \psi^{8} -\frac{r}{10!}\psi^{10}+ \ldots,
\end{equation}

for some branch of the logarithm. 

The logarithm in the exact representation may be understood pointwise with an arbitrary branch, since the branch ambiguity drops out after exponentiation. At zeros of \(\tilde V\), the integrand is understood directly through the equivalent product representation. The perturbative expansion below, however, uses a single analytic branch of \(\log\tilde V\) in a neighborhood of the origin.

Since \(\tilde V(0)>0\), there exists a neighborhood of \(\psi=0\) in which \(\tilde V(\psi)\neq0\) and an analytic branch of \(\ln\tilde V(\psi)\) can be chosen. Different choices of the logarithm branch in this neighborhood differ only by an additive constant \(2\pi i k\) and therefore give the same nonconstant coefficients in the expansion above. In what follows, we use this expansion only as a local power-series representation around \(\psi=0\); no global representation of \(\ln\tilde V\) across the zeros of \(\tilde V\) is assumed. In the dual strong-coupling expansion, the argument of this function is \(\psi/g\), so the expansion naturally generates powers of \(1/g\).

The explicit formulas for the expansion coefficients $a$, $b$, $c$, $d$ can be found in Appendix~\ref{appendix:dual-action-coeffs}. For the case considered below $n=2$ ($\phi^4$ theory) they are numerically equal to:

\begin{equation}
    a \approx 1.65580, \quad b \approx 2.22504, \quad c \approx 16.9726, \quad d \approx 288.362, \quad r \approx 8497.59,
\end{equation}

and this precision will be sufficient for the numerical simulations and calculations as confirmed below.

Substituting \eqref{HS-duality} into \eqref{PartFunc1} for every site:

\begin{equation}
    \exp\left(- \sum\limits_x \frac{g^{2n}\phi
    (x)^{2n}}{(2n)!} \right) = \frac{1}{(2\pi g)^N} \int\limits_{\mathbb{R}^N} \prod\limits_x d\psi(x) \ e^{\sum\limits_x \ln \tilde{V}(\psi(x)/g) + i \sum\limits_x \phi (x) \psi(x) },
\end{equation}

we can take the Gaussian integrals over $\phi$, and perform the duality transformation, writing down the dual expression for partition function:

\begin{equation}
\mathcal{Z}[j(x)] =\frac{1}{g^N}\int_{\mathbb{R}^{N}}\prod_{x}\frac{d\psi(x)}{2\pi}\exp\left(-\frac{1}{2}\sum \limits_{x,x^{'}} \left(\psi(x)-j(x)\right)G_{x,x^{'}} \left(\psi(x^{'})-j(x^{'})\right)+\sum_{x}\ln \tilde{V}\left( \frac{\psi}{g} \right)\right).
\end{equation}

Let us not that the integrals over fields converge absolutely for $g>0$ because $|e^{\ln\,V(\psi/g)}|\leq \tilde{V}(0)$. Introducing the dual action:

\begin{equation}
\label{dual-action-expanded}
S^{*}[\psi]=\frac{1}{2}\sum \limits_{x,x^{'}} \left(\psi(x)-j(x)\right)G_{x,x^{'}} \left(\psi(x^{'})-j(x^{'})\right)+\sum\limits_{x} \left( \frac{a}{2} \frac{\psi^{2}(x)}{g^2}+\frac{b}{24} \frac{\psi^{4}(x)}{g^4}+\frac{c}{6!} \frac{\psi^{6}(x)}{g^6} + \ldots \right),
\end{equation}
and the strong-coupling ``background`` energy $\mathcal{F}_0$ as:

\begin{equation}
\label{F0-strong-def}
\frac{\tilde{V}(0)^{N}}{(2\pi)^{N/2}g^N\sqrt{\det \left( G + \frac{a}{g^2} \right)}}=\exp(-\tilde{\mathcal{F}}_{0}),
\end{equation}

one can rewrite the partition function in the form:

\begin{equation}
    \label{dual-partition-general-final}
\mathcal{Z}[j(x)] =e^{-\tilde{\mathcal{F}}_0}\int_{\mathbb{R}^{N}}\frac{\prod_{x}d\psi(x)}{(2\pi)^{N/2}\sqrt{\det \left( G + \frac{a}{g^2} \right)^{-1}}}\exp\left(-S^*[\psi] \right).
\end{equation}

In the equation above we have deliberately added the normalization factor to the integration measure, corresponding to the dual action quadratic terms structure. The prefactor here corresponds to the strong-coupling background free energy. Let us denote the corresponding energy density as:

\begin{equation}
\label{f0-strong-def}
    \tilde{f}_0 = \lim\limits_{N\rightarrow\infty} \frac{\mathcal{\tilde{F}}_0}{N}.
\end{equation}

In Appendix~\ref{appendix:strong-coupling-bg-free-en}, it is shown that for $N\gg 1$ there exists a finite limit for $\tilde{f}_0$:

\begin{equation}
\label{f0-strong}
    \tilde{f}_0 =  \ln \frac{\sqrt{2\pi}}{\tilde{V}(0)} + \frac{1}{2 (2\pi)^{d}} \left(\prod_{j=1}^d \int_{0}^{2 \pi} dq_{j}\right)\ln\frac{g^2 + a\left(4\alpha\sum_{j=1}^{d}\sin^{2}\left(\frac{q_{j}}{2}\right)+\gamma\right)}{4\alpha\sum_{j=1}^{d}\sin^{2}\left(\frac{q_{j}}{2}\right)+\gamma}.
\end{equation}

It will be also convenient to eliminate the unnecessary quantities from the dual action, such as constant terms and the contributions of the source $j(x)$. For this purpose, let us define the reduced action of a dual theory, $S_0^*[\psi]$ as:

\begin{equation}
\label{dual-action-shortened-expanded}
S_0^{*}[\psi]=\frac{1}{2}\sum \limits_{x,x^{\prime}} \psi(x) \left(G_{x,x^{\prime}}+\frac{a}{g^2}\delta_{x,x^{\prime}}\right) \psi(x^{\prime})+\sum\limits_{x} \left(\frac{b}{24} \frac{\psi^{4}(x)}{g^4}+\frac{c}{6!} \frac{\psi^{6}(x)}{g^6} + \frac{d}{8!}\frac{\psi^{8}(x)}{g^8} + \frac{r}{10!}\frac{\psi^{10}(x)}{g^{10}}+\ldots \right),
\end{equation}

where we deliberately distinguished quadratic and higher order terms in action. Now it is useful to introduce the Green's function of the dual theory in the strong-coupling expansion:

\begin{equation}
\label{dual-theory-propagator-def}
    \tilde{G}^{-1}=\left(G_{x,x^{\prime}}+\frac{a}{g^2}\delta_{x,x^{\prime}}\right),
\end{equation}

being the inverse of the quadratic part operator of the dual action \eqref{dual-action-shortened-expanded}. In momentum space, it takes the following form (see Appendix~\ref{appendix:strong-coupling-gf-calc}):
\begin{equation}
\label{dual-theory-propagator}
     \tilde{G}(q) = \frac{g^2}{a}\frac{4\alpha \sum_{j=1}^{d}\sin^{2}\left(\frac{q_{j}}{2}\right)+\gamma}{4\alpha\sum_{j=1}^{d}\sin^{2}\left(\frac{q_{j}}{2}\right)+\left(\frac{g^{2}}{a}+\gamma\right)}.
\end{equation}

Moreover, for later use it will be helpful to define also the partition function of the dual theory in the external field $\eta(x)$:

\begin{equation}
\label{dual-partition}
    \mathcal{Z}^*[\eta(x)] = \int_{\mathbb{R}^{N}}\frac{\prod_{x}d\psi(x)}{\sqrt{(2\pi)^ N \det \tilde{G}}} \exp\left[ -S_0^*[\psi] +  i \left \langle \eta(x), \psi(x)\right \rangle\right],
\end{equation}

where we have used the normalization factor imposed by the quadratic part of the dual action $S^*$. This formula could be considered as a definition of a dual theory. Crucially, one can compute the correlators in such a theory, using the standard Feynman-diagram expansion, and their finite-order truncations approximate much better than in the original theory \eqref{lattice-action}. So let us define the correlators in dual theory as:

\begin{equation}
    \left \langle \psi(x_1) \cdot \ldots \cdot \psi(x_k) \right \rangle^* = \frac{1}{\mathcal{Z}^*[0]} \int_{\mathbb{R}^{N}}\frac{\prod_{x}d\psi(x)}{\sqrt{(2\pi)^ N \det \tilde{G}}} \ \psi(x_1) \cdot \ldots \cdot \psi(x_k) \ \exp \left(-S_0^*[\psi]\right).
\end{equation}

Clearly, such a general correlation can be expressed in terms of functional derivatives of $\mathcal{Z}^*[\eta(x)]$ with respect to $\eta(x)$ in a common way.

Proceeding, one can also express the correlators of initial theory in terms of the correlators of dual theory. It can be seen from the relation between the partition functions of the initial and dual theories, which reads as:

\begin{equation}
    \mathcal{Z}[j(x)] = \exp \left[ -\widetilde{\mathcal{F}}_0 - \frac{1}{2}\sum_{x,x^{'}}j (x) G_{x,x'}j(x')\right] \mathcal{Z}^*[-i\cdot(Gj)(x)],
\end{equation}

where $(Gj)(x)=\sum_{x'} G_{x,x'} j(x')$.

This relation allows us to deduce that:

\begin{equation}
\label{correlators-link}
     \langle \phi(p) \phi(-p) \rangle = G(p) \left( 1 - G(p) \left \langle \psi(p) \psi(-p) \right \rangle^* \right),
\end{equation}

where $\langle \phi(p) \phi(-p)  \rangle$ is a Discrete Fourier Transform \eqref{DFT} of a correlation function $\left \langle \phi(x) \phi(0) \right \rangle$ \eqref{correlators-def-initial}, and similarly for the dual theory correlator $\langle \psi(p) \psi(-p) \rangle^*$. The expression above connects the two-point correlation functions in the initial and dual theories, and can be directly extended to correlators of higher orders. We will use this relation for calculation of the two-point correlation function in the strong coupling expansion in the Section~\ref{sect:two-point-func}.

Let us briefly discuss the obtained result. The initial action $S$ was the simple action of $\phi^{2n}$ theory on the lattice, where $g$ is a coupling constant and $\phi$ is a field variable. It provides a common Feynman expansion in powers of $g$. Then we applied the constructed duality and got the interaction in a dual action $S^*$ as a power series in $\psi / g$, where $\psi$ is a new (dual) field variable. As a result, the (lattice) path integral with a dual action $S^*$ can provide a Feynman perturbation theory in the inverse powers of coupling constant $g$. The obtained dual theory action is complex valued. Hence, the constructed duality serves as a roadmap for calculating strong coupling expansions at least for power potentials. 

Let us also highlight the fact that after applying the duality some ``mass renormalization`` occurred, namely, we have obtained some nonzero  contribution to the quadratic part of action from $\ln \tilde{V}$. That is why we normalized integrations on $\det \left( G + \frac{a}{g^2} \right)^{-1}$, rather than only $\det G^{-1}=\det L$. Though even the notion of a mass is not clear in the dual theory, since it is essentially non-local. The propagator also becomes dependent on the coupling constant. 

Now, let us present the results of duality application to the case of free energy density.

\section{Free energy density calculation}
\label{sect:free-energy}

In this section, we apply the described duality to the calculation of the free energy per site \eqref{free-energy-per-site-def} in thermodynamic limit. We have chosen this quantity to show an example of calculations since it has no dependence on the number of lattice sites and the boundaries.  We also compare strong and weak coupling expansions with each other and numerical simulations. 

From this moment, we will consider only the case of $n=2$,  which corresponds to $\phi^4$ theory.

For both strong- and weak-coupling cases one can derive the expressions for free energy per site using common Feynman Perturbation Theory. The detailed calculations can be found in Appendix~\ref{appendix:pert-free-energy-calc} .  To verify the obtained results, we use the numerical simulations with Hamiltonian Monte Carlo Method \cite{HMC_START_DUANE,Modern_Perspectives_in_Lattice_QCD,neal2012mcmc,PhysRevD.96.114505_lattice_hmc_simulations_1, Pawlowski_2020_lattice_hmc_simulations_2} for cubic lattice with $M=32$ sites in every dimension and periodic boundary conditions.  Thorough specifications for the performed numerical simulations can be found in Appendix~\ref{appendix:num-methods} .

For the strong-coupling case one arrives at the expression, decomposing \eqref{dual-partition-general-final}:

\begin{figure}
    \centering
    \includegraphics[width=8cm,height=5.5cm]{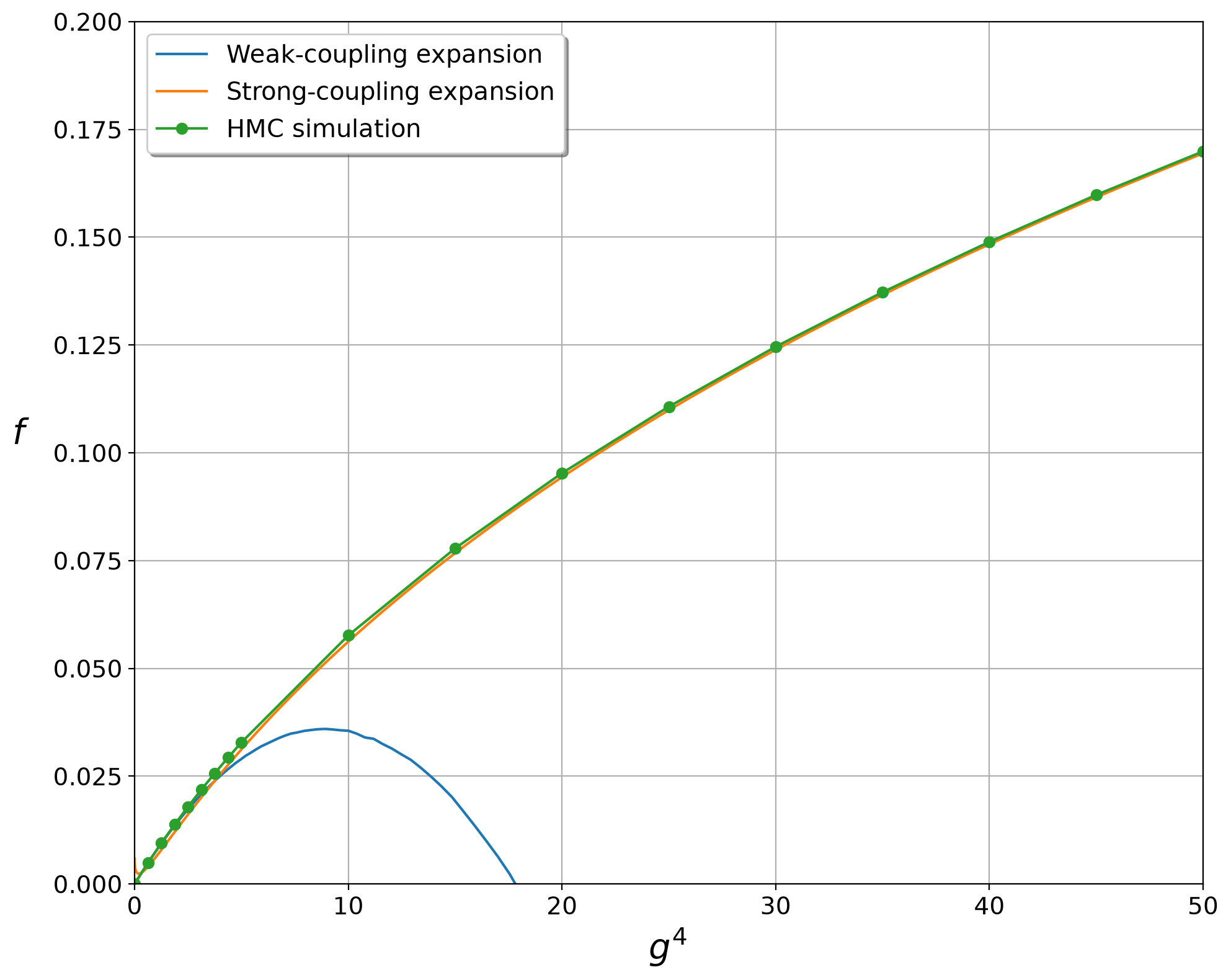}
    \hfill
    \includegraphics[width=8cm,height=5.5cm]{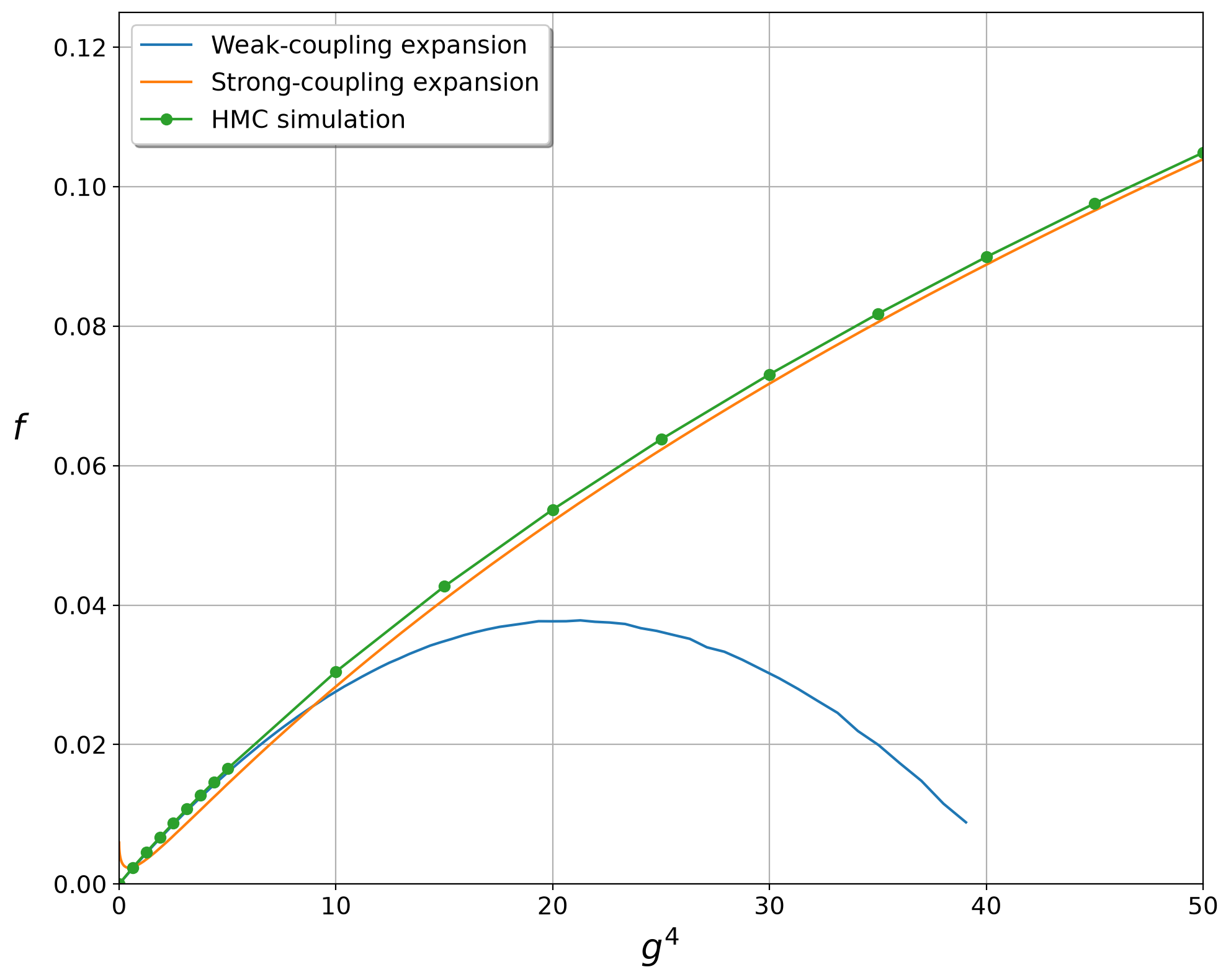}
    \caption{Plots of analytical and computational results for the free energy per site \eqref{free-energy-per-site-def} with normalization with respect to free theory. The left picture corresponds to the dimension $d=2$ and the right one to the dimension $d=3$. As the dimension increases, the obtained strong coupling expansion seems to approach numerical curve more slowly.}
    \label{fig:f(g)-comparison}
\end{figure}
\begin{equation}
\label{free-energy-strong_main}
    \begin{split}
 f &=\tilde{f}_0+\frac{b}{8 g^{4}}(\tilde{G}_0)^{2}+\frac{c}{2^{3}3! g^{6}}(\tilde{G}_0)^{3}-\frac{b^{2}}{2^{4}(2 \pi)^{d} g^{8}}\left((\tilde{G}_0)^{2}\left(\prod_{i=1}^{d}\int_{0}^{2\pi}dq_{i}\right)\tilde{G}(q)^{2}\right)\\
 &+ \frac{d}{2^4 4! g^8} \left(\tilde{G}_0\right)^4 -\frac{b^{2}}{48 (2 \pi)^{3d}  g^{8}}\left(\prod_{n=1}^{3}\left(\prod_{i=1}^{d}\int_{0}^{2\pi}dq_{i}^{n}\right)\tilde{G}\left(q^{n}\right)\right)\tilde{G}\left(q^{1}+q^{2}+q^{3}\right) + O\left(1/g^{10}\right),
    \end{split}
\end{equation}

Let us analyze the expression obtained. At first glance, the series diverges in the limit $g\to0$. However, in this limit the dual Green's function $\tilde{G}\propto g^2$, and the inverse powers of the coupling constant from the expansion of the potential are canceled by the positive powers from the Green's function, preserving regularity at $g\to0$.  Unlike many other strong expansions \cite{strong_coupling_Bender,Kleinert:1998-strong-coupling,FRASCA_2007_PT_duality,Ising-low-t-decompositions-I-G-Enting-1994}, in the presented one all the terms are regular when $g \rightarrow 0$. This regularity does not imply that the strong-coupling truncation is accurate in the weak-coupling regime, but allows to avoid the accuracy affect caused by expansion terms singularity near zero.

Another notable feature of the given series is the leading order of the free energy in the strong-coupling regime grows logarithmically: $\tilde{f}_0\propto\ln g$ for $g \to \infty$, which is derived in Appendix~\ref{appendix:strong-coupling-bg-free-en} .

The comparative plots of $f(g)$ for $\alpha=1$ and $\gamma=1$ can be found in Figure~\ref{fig:f(g)-comparison}  for $d=2$ and $d=3$ . For the sake of verification, we have also placed standard weak coupling series for free energy per site (Appendix~\ref{appendix:pert-free-energy-calc}) on the same plots. One can see that weak series agrees with numerical simulations at small coupling, and strong at intermediate and large coupling with overlapping domains of application. It indicates that both expansions are consistent with each other. In the process, the weak perturbation series obtained coincides with numerical simulation in the region of small couplings $g$. Finally, the obtained strong coupling expansion \eqref{free-energy-strong_main} is placed close to the numerical simulations, validating the approximation over the tested parameter range. 

The obtained strong coupling fits the numerical results sufficiently well in the region of medium and large $g$. 

For $d=2$, the relative deviation of the strong-coupling approximation from the HMC result decreases to approximately $5\%$ by $g^4\simeq5$, to about $2\%$ at $g^4=10$, and below $1\%$ for sufficiently strong coupling. In $d=3$, the approach is slower: the deviation falls below $5\%$ around $g^4\simeq15$ and reaches the percent level only at substantially larger coupling. For details see Appendix~\ref{appendix:num-methods}.

This provides an independent consistency check for an integrated observable of the provided technique. We also want to investigate the coordinate (or momentum) dependence of this perturbation series. So, we are going to study correlators, restricting ourselves to the case of two-point correlation functions.

\section{Two-point correlation function}
\label{sect:two-point-func}

While the free-energy density, Eq.~\eqref{free-energy-per-site-def}, provides an integrated characterization of the theory, a more stringent test of the dual expansion is whether it also reproduces nontrivial spatial dependence. We therefore consider the two-point correlation function, or equivalently its momentum-space representation. The simplest, yet still useful, example is the two-point function \eqref{two-point-momentum-def} generated by $\mathcal{Z}[j(x)]$ \eqref{PartFunc1}.

\begin{figure}
    \centering
    \includegraphics[width=8cm,height=5.5cm]{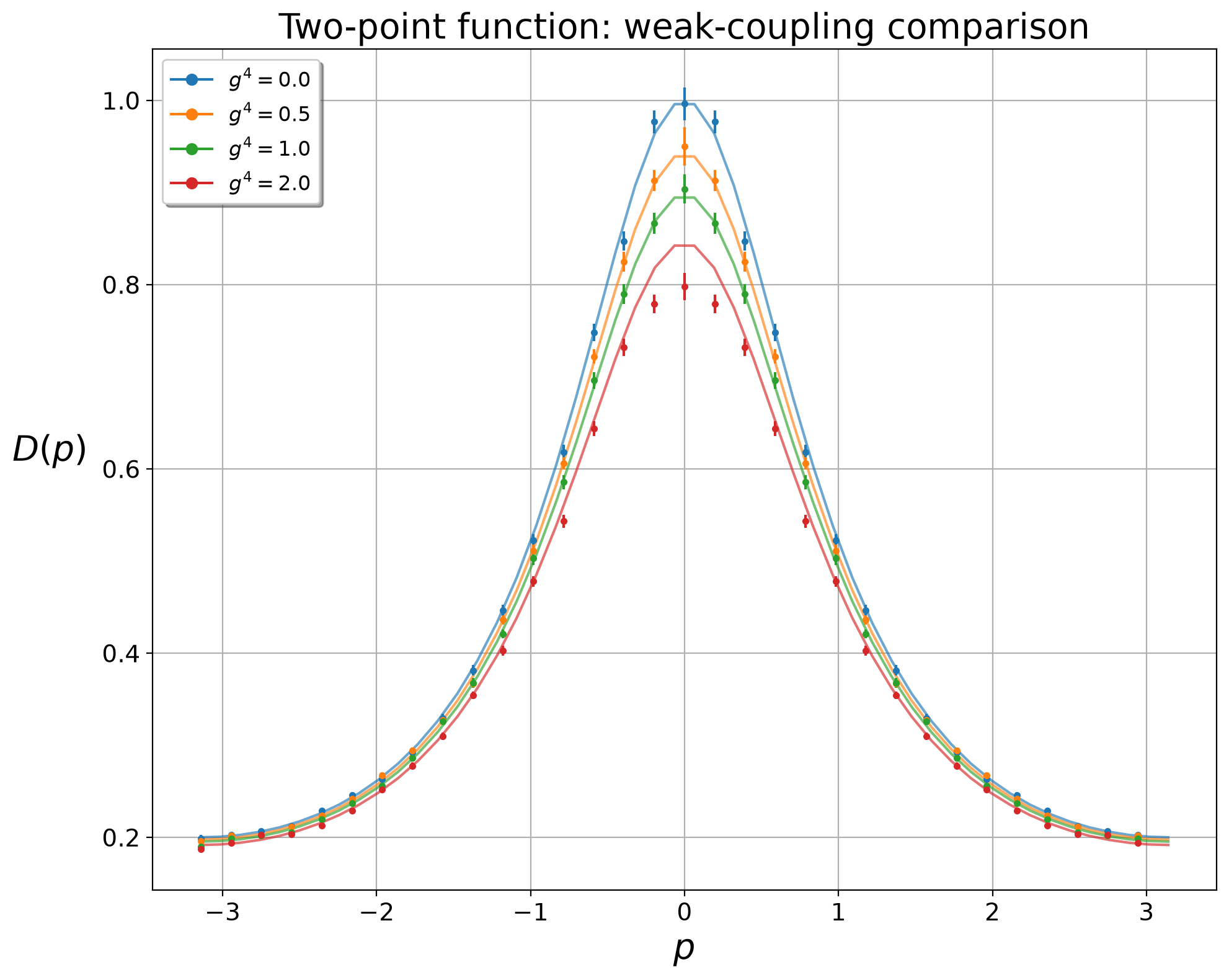}
    \hfill
    \includegraphics[width=8cm,height=5.5cm]{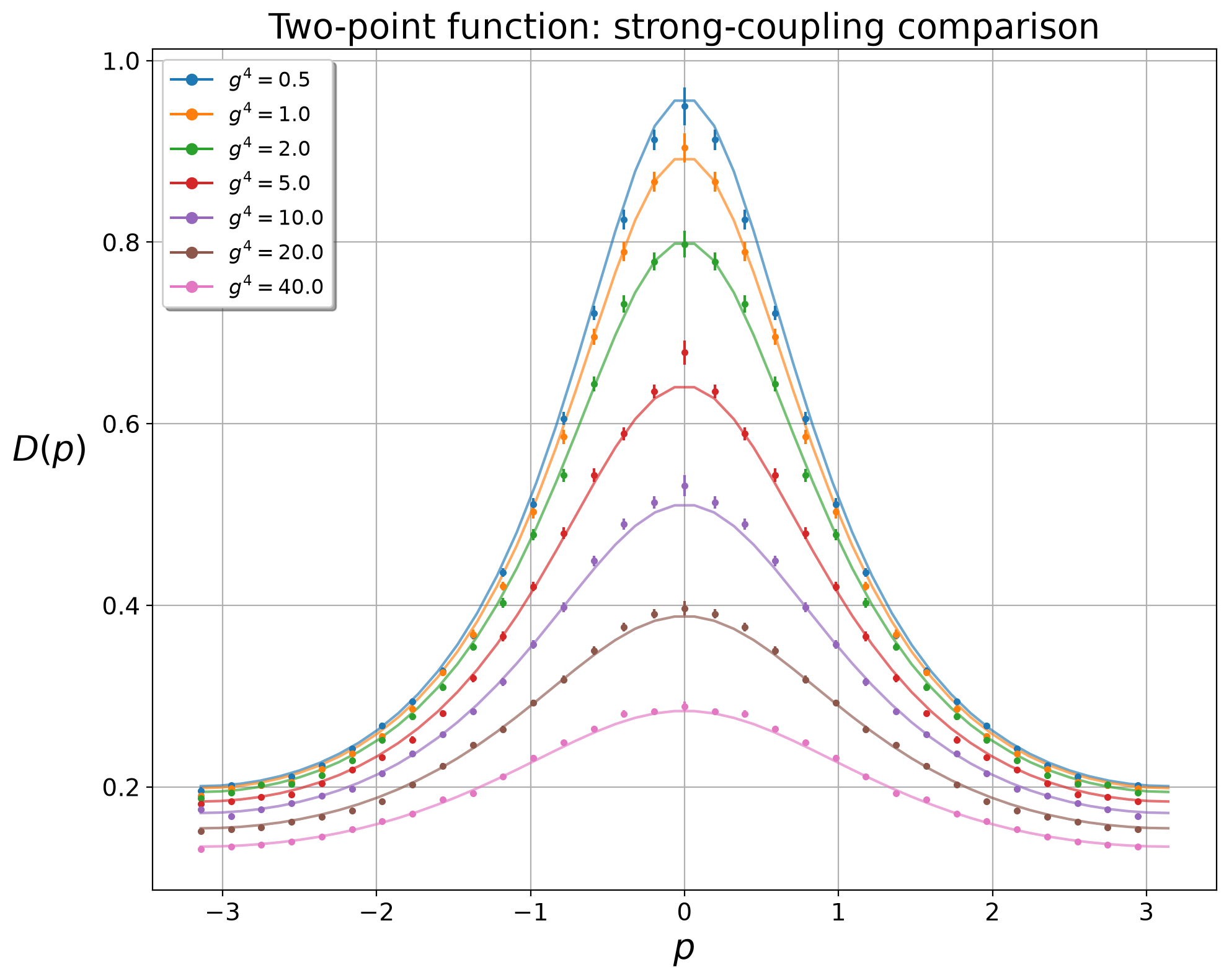}
    
    \includegraphics[width=8cm,height=5.5cm]{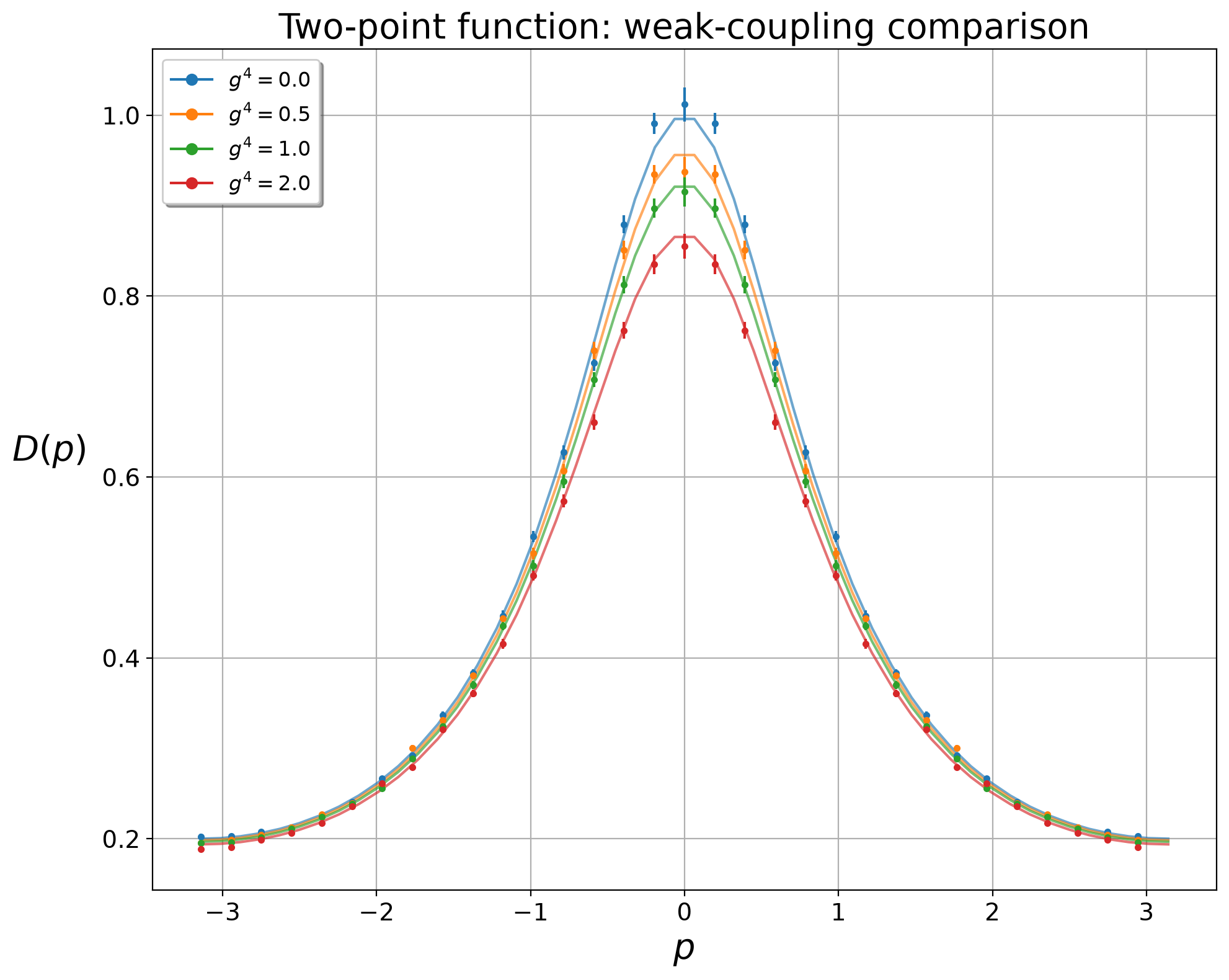}
    \hfill
    \includegraphics[width=8cm,height=5.5cm]{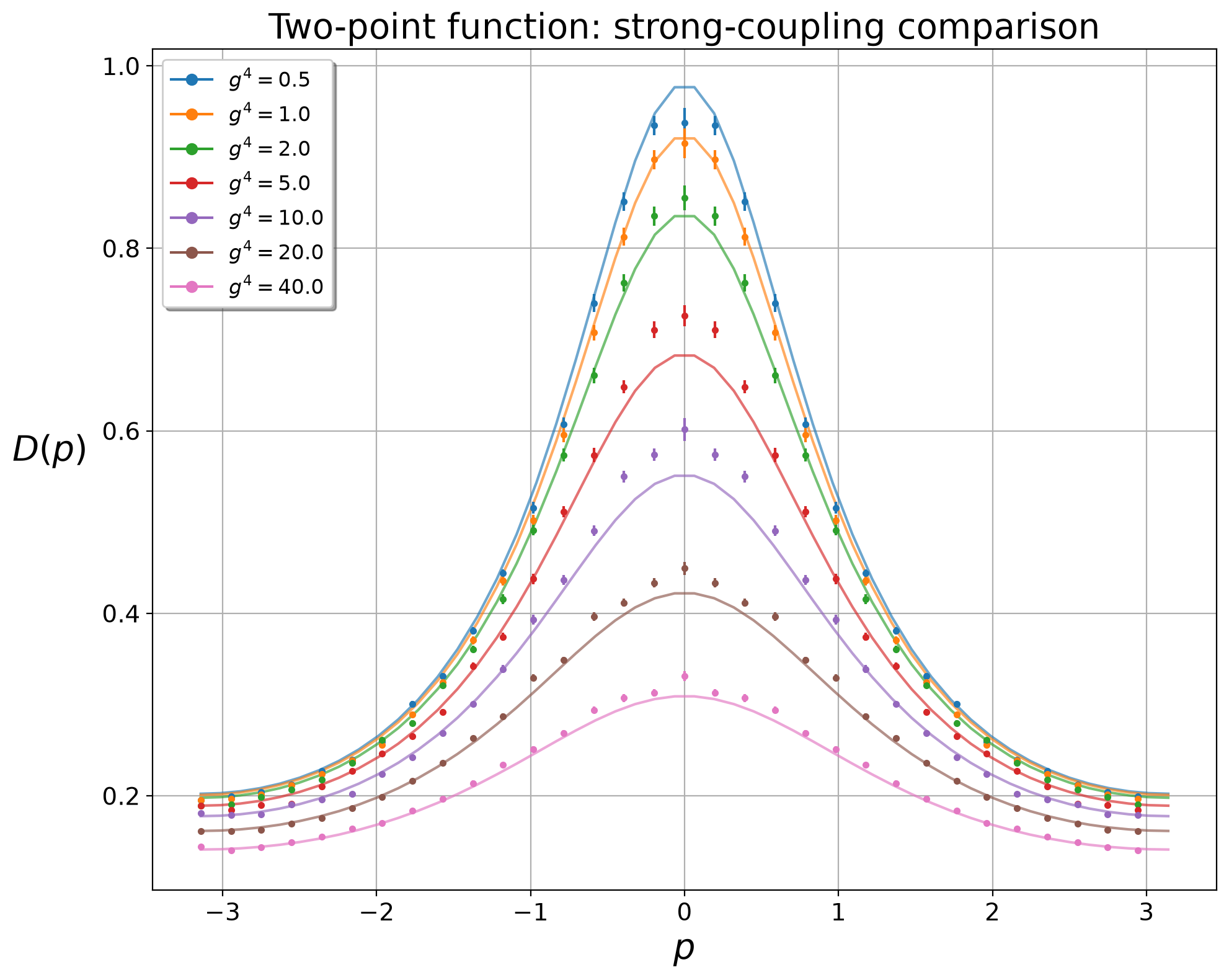}
    \caption{Momentum-space two-point function $D(p)=\langle \phi(p)\phi(-p) \rangle$ \eqref{two-point-momentum-def} comparison for $d=2$ (top) and $d=3$ (bottom) for weak \eqref{two-point-weak} and strong \eqref{two-point-strong} coupling expansions. Here vector momentum is parametrized as $(p,0)$ for $d=2$ and $(p,0,0)$ for $d=3$. The dots with error bars correspond to numerical simulations with Hamiltonian Monte Carlo described in Appendix~\ref{appendix:num-methods}. The solid lines denote the computations with \eqref{two-point-weak} and \eqref{two-point-strong} as indicated by the labels. Though strong coupling approximation was anticipated, the reason for the \eqref{two-point-strong} to give admissible results for intermediate couplings is a remarkable feature, following from the regularity of this expansion and the fact that for $g=0$ the relation \eqref{correlators-link} degenerates to $G(p)$. The traditional weak coupling perturbation theory fails for $g^4 \geq 2$, which could be expected from their asymptotic nature.}
    \label{fig:two-point-comparison}
\end{figure}

To obtain the correlation function of the original theory, we have to link it to the dual theory using \eqref{correlators-link}. As in the case of the free energy, one can calculate the two-point function using Feynman diagrams (see Appendix~\ref{appendix:two-point}). In the present section we write down only the final answer.

Let us introduce the notations for momentum integrals as:

\begin{equation}
\label{integrals-two-point}
\begin{split}    
    \tilde{I}_2 &= \frac{1}{(2 \pi)^d}\left(\prod_{i=1}^{d}\int_{0}^{2\pi}dq_{i}\right) \tilde{G}(q)^2, \\
    \tilde{I}_3(p) &= \frac{1}{(2\pi)^{2d}} \left(\prod_{i=1}^{d}\int_{0}^{2\pi}dq_{i}\right)\left(\prod_{i=1}^{d}\int_{0}^{2\pi}dk_{i}\right) \tilde{G}(k) \tilde{G}(q) \tilde{G}(q+k + p), \\
    \tilde{I}_4 &= \frac{1}{(2\pi)^{3d}} \left(\prod_{n=1}^{3}\left(\prod_{i=1}^{d}\int_{0}^{2\pi}dq_{i}^{n}\right)\tilde{G}\left(q^{n}\right)\right)\tilde{G}\left(q^{1}+q^{2}+q^{3}\right).
\end{split}
\end{equation}

Then the final formula reads:

\begin{equation}
\label{two-point-strong}
\begin{split}
\frac{\left\langle \psi(p)\psi(-p)\right\rangle^* }{\tilde{G}(p)^2} & =\frac{1}{\tilde{G}(p)}-\frac{b}{2 g^4}\tilde{G}_0 - \frac{c}{2^3 g^6} \left(\tilde{G}_0\right)^2 - \frac{d}{3! 2^3 g^{8}}\left( \tilde{G}_0 \right)^3 + \frac{b^2}{2^2 g^8} \tilde{G}_0 \tilde{I}_2 +\frac{b^2}{2^2 g^8} \left(\tilde{G}_0\right)^2 \tilde{G}(p) \\ & + \frac{b^2}{3! g^8 } \tilde{I}_3 (p) -\frac{r}{4!2^4g^{10}}\left(\tilde{G}_0\right)^4 + \frac{bc}{8 g^{10}} \left(\tilde{G}_0\right)^3 \tilde{G}(p) + \left( \frac{1}{8} + \frac{1}{16} \right) \frac{bc}{g^{10}} \left(\tilde{G}_0\right)^2\ \tilde{I}_2
 \\& +\frac{bc}{6g^{10}} \tilde{G}_0 \tilde{I}_3(p) +\frac{bc}{24 g^{10}} \tilde{I_4} + O\left(\frac{1}{g^{12}}\right) , \qquad g \rightarrow \infty
\end{split}
\end{equation}
where we define $\tilde{G}_0=\tilde{G}(x=0)$ as the dual theory Green's function \eqref{dual-theory-propagator} at zero coordinate.

Let us note that for $g\rightarrow0$ the expression obtained reads:

\begin{equation}
    \left\langle \psi(p)\psi(-p)\right\rangle^* \sim g^2,
\end{equation}

being regular for small couplings. In $d=2$, the $g^{-10}$ strong-coupling truncation shows good agreement with the HMC data over the tested intermediate- and strong-coupling range. In $d=3$, approximation is worse and a residual systematic deviation remains, although the agreement improves toward stronger coupling. This residual deviation indicates that higher-order corrections remain relevant in this case.

From the plots in Figure~\ref{fig:two-point-comparison} one can deduce that the results of numerical simulations are in agreement with the standard weak-coupling perturbation theory \eqref{two-point-weak} at sufficiently small coupling (Appendix~\ref{appendix:two-point}).

These results indicate that the dual expansion can provide quantitatively useful finite-order approximations for lattice observables in the model studied here. The largest deviations occur in the infrared region near $p=0$, while the overall momentum dependence is reproduced more accurately.

\section{Discussion}
\label{sect:discussion}

In this work, we developed a strong-coupling expansion based on a field-space Fourier representation of scalar lattice field theory and applied it to the lattice $\phi^4$ model.  All the expansions are obtained using standard Feynman diagram techniques. Our analysis focused on the free energy per site $f$ \eqref{free-energy-per-site-def} and the two-point correlation function $\left< \phi(p) \phi(-p) \right>$ in momentum space \eqref{two-point-momentum-def} for dimensions $d=2$ and $d=3$. In the strong- and intermediate-coupling regimes where the expansion is expected to be applicable, the finite-order analytical results show good agreement with independent numerical simulations. Quantitatively, the strong-coupling approximation of free energy per site reaches few-percent accuracy already at intermediate coupling, although the onset of this regime is noticeably slower in $d=3$ than in $d=2$. For the two-point function, the agreement is weakest at low momentum and improves progressively toward larger $p$.

We have considered these quantites for the following reasons. The free-energy calculation tests the ability of the expansion to reproduce an observable obtained after integrating over the lattice, whereas the two-point function provides a more sensitive check for a quantity with nontrivial momentum or coordinate dependence. The agreement observed for the latter indicates that the dual expansion captures not only bulk quantities but also the spatial structure encoded in correlation functions.

Several qualifications are worth emphasizing.

First, the present analysis concerns lattice theories at finite lattice spacing. Whether the construction admits a useful continuum formulation, and how it relates to renormalization-group properties such as fixed points or quantum triviality, are left for the future work.

Second, the dual action derived here may also provide a starting point calculations other than the perturbative expansion considered in the present work. It would be interesting to investigate if saddle-point or other nonperturbative techniques applied directly in the dual variables lead to useful results.

Finally, the numerical results suggest that the strong-coupling expansion can remain quantitatively useful beyond the asymptotically large-coupling regime. We therefore view the strong-coupling limit as the controlled starting point of the expansion, while its practical range of applicability may extend into intermediate coupling depending on the quantity considered.

\section{Conclusions}
\label{sect:conclusions}

The field-space Fourier representation developed in the present work provides a systematic diagrammatic strong-coupling expansion for scalar lattice theories.

The construction is not specific to the quartic interaction.It can be applied more generally to bosonic lattice theories with site-local interactions, supposing they admit a suitable Fourier transform and that their transform is nonzero and analytic in a neighborhood of the expansion point. Specifically, these include real scalar fields with a general interaction $V(\phi)$ and complex scalar fields with a general interaction $V(\phi, \phi^*)$.

Further numerical studies in higher dimensions would be useful. It would also be interesting to investigate alternative approximation schemes directly in the dual variables, including saddle-point methods, and to study systematically the behavior of the dual formulation in the approach to the continuum limit. Extensions to lattice gauge theories or theories with fermionic degrees of freedom would require additional structure and are left for future work.

Another possible direction concerns the analytic structure of correlation functions obtained from the strong-coupling representation. After taking the thermodynamic limit and performing an appropriate analytic continuation, one could investigate the resulting poles, branch points, and branch cuts and their relation to the spectral structure of the theory. Establishing such a correspondence would require an analysis beyond the finite-order calculations presented here.

\section*{Acknowledgements}

The authors are grateful to V. Kiselev, A. Litvinov, E. Akhmedov, S. Ogarkov and M. Lashkevich for helpful discussions and criticism.

\section*{Funding information}
The authors received no specific funding for this work.

\printbibliography

\begin{appendices}

\section{Derivation of discrete Laplacian eigenvalues}
\label{appendix:laplacian}

In this appendix, we are going to discuss in
more detail, how eigenvalues of lattice Laplacian depend on grid
scale $l$. More examples of computing the Laplacian spectrum for various lattices and boundary conditions can be found, for example, in \cite{networks-graphs-pozrikidis2014}. 

First of all, we address the dimensionless problem. Let us consider a graph which is a $d$-dimensional hypercubic lattice with ``unit`` spacing with $M$ sites in every direction and periodic boundary conditions. Let us also denote  the total number of sites or, equivalently, modes, as $N=M^{d}$. To underline the vector nature of the performed additions, we will use the notation $\vec{r}$ for the lattice site position and the general function on lattice as $f(\vec{r})$. Further, one can easily return the lattice spacing to formulas using dimensional analysis. So let us define the lattice Laplacian on such lattice as:

\begin{equation}
\label{lattice-laplacian-def-appendix}
(\triangle f)(\vec{r})=\sum_{j=1}^{d}\left(f(\vec{r}+\vec{e}_{j})-2f(\vec{r})+f(\vec{r}-\vec{e}_{j})\right),
\end{equation}
where $\vec{e}_{j}$ is the $j$th vector of an orthonormal basis, corresponding
to the cubic lattice under consideration. As one can see, for the lattice Laplacian under consideration
it is true that:
\begin{equation}
\triangle=-D+J,
\end{equation}
where:
\begin{enumerate}
\item $D$ is a scalar matrix with elements enumerated by sites and
equal to the degree of the corresponding site, i.e., the number of edges attached
to it. So, in $d$-dimensional cubic lattice $D_{\vec{r},\vec{r}'}=2d\delta_{\vec{r},\vec{r}'}$. 
\item $J$ is an adjacency matrix of the grid graph under consideration, i.e. cubic
lattice. Evidently, for a cubic lattice, its matrix elements are given by: 
\begin{equation}
\label{adjacency-matrix-cubic}
    J_{\vec{r},\vec{r}'}=\sum_{j=1}^{d}\left(\delta_{\vec{r}^\prime,\vec{r}+\vec{e}_{j}}+\delta_{\vec{r}^\prime,\vec{r}-\vec{e}_{j}}\right)
\end{equation}

\end{enumerate}

Therefore, finding the spectrum of the Lattice Laplacian $\Delta$ is equivalent to finding the spectrum of the corresponding lattice graph which is, by definition, the spectrum of its adjacency matrix \eqref{adjacency-matrix-cubic}.

So, let us diagonalize $J$. We are going
to find the complete set of lattice eigenfunctions. The corresponding
equation for eigenfunctions:
\begin{equation}
\sum_{\vec{r}'}J_{\vec{r},\vec{r}'}f(\vec{r}')=\mu f(\vec{r}),\qquad\sum_{j=1}^{d}(f(\vec{r}+\vec{e}_{j})+f(\vec{r}-\vec{e}_{j}))=\mu f(\vec{r}),
\end{equation}
which is no more than a linear homogeneous difference equation. Its eigenmodes can be chosen as plane waves (for general parameters), namely in the form of
exponents of linear functions of indices, i.e. $e^{i(\vec{q},\vec{r})}$
in our vector notations. This form of solution is also clear from
physical intuition, since it is a plane wave solution. Hence, one
can get an equation, substituting $f(\vec{r})=e^{i(\vec{q},\vec{r})}$:
\begin{equation}
\sum_{j=1}^{d}(e^{i(\vec{q},\vec{e}_{j})}+e^{-i(\vec{q},\vec{e}_{j})})=\mu,\qquad\mu=2\sum_{j=1}^{d}\cos q_{j},
\end{equation}
where $\vec{q}=\sum_{j=1}^{d}q_{j}\vec{e}_{j}$. From boundary conditions:
\begin{equation}
f(\vec{r}+M\vec{e}_{j})=f(\vec{r}),
\end{equation}

we find that:

\begin{equation}
    q_{j}^{(k)}=\frac{2\pi k_{j}}{M},\quad k_{j}=0,1,\ldots,M-1
\end{equation}

Therefore, we have found $M^{d}$ eigenvalues counted with multiplicity, 
diagonalized $J$ and hence $\Delta$. As a result, the eigenvalues
of discrete lattice Laplacian are $\left( \vec{q}=\sum_{j=1}^{d}q_{j}\vec{e}_{j}\right)$:
\begin{equation}
\label{laplacian-eigenvalues-discrete-final-appendix}
\lambda_{\vec{q}}=-2d+2\sum_{j=1}^{d}\cos q_{j},\quad q_{j}=\frac{2\pi k_{j}}{M},\ k_{j}=0,1,\ldots,M-1
\end{equation}

Recovering the lattice spacing $l$ from the dimensions of parameters, we get:
\begin{equation}
\lambda_{\vec{q}}=-\frac{2}{l^{2}}\sum_{j=1}^{d}\left(1-\cos q_{j}\right),\qquad q_{j}=\frac{2\pi k_{j}}{M},\qquad k_{j}=0,\ldots,M-1.
\end{equation}

In this paper we don't consider the continuum limit, so we should stop at the expression \eqref{laplacian-eigenvalues-discrete-final-appendix}.

Let us add a few words about the continuum limit of the obtained eigenvalues to validate the obtained answer. To perform the continuum limit, we would like to change variables to momenta $\vec{p}=\frac{1}{l}\vec{q}$ from the lattice quantum numbers $q_j$, since they are the real
physical quantities which are reasonable to keep finite. So, let us substitute $\vec{p}=\frac{1}{l}\vec{q}$, and then, simplifying, we obtain:
\begin{equation}
\lambda_{\vec{p}}=-\frac{4}{l^{2}}\sum_{j=1}^{d}\sin^{2}\left(\frac{p_{j}l}{2}\right),
\end{equation}
where:
\begin{equation}
\vec{p}=\sum_{j=1}^{d}p_{j}\vec{e}_{j},\quad p_{j}=\frac{2\pi k_{j}}{Ml},\quad k_{j}=0,1,\ldots,M-1,
\end{equation}
so, we get for $l\rightarrow0$, $\vec{p}=\text{const}$ and
$M\cdot l=\text{const}$: 
\begin{equation}
\lambda_{\vec{p}}=-\frac{4}{l^{2}}\sum_{j=1}^{d}\sin^{2}\left(\frac{p_{j}l}{2}\right)\rightarrow\lambda_{\vec{p}}^{(c)}=-\sum_{j=1}^{d}p_{j}^{2}=-\vec{p}^{2},
\end{equation}
and momenta values are not bounded now, after continuum limit. So, we have obtained the continuum limit of eigenvalues $\lambda_{\vec{p}}^{(c)}$, which coincides with the spectrum of continuous Laplacian on a (periodic) hypercube of dimension $d$, which verifies the obtained discrete answer.

\section{Explicit formulas for dual action expansion coefficients }
\label{appendix:dual-action-coeffs}
In this appendix, we provide explicit formulas for dual interaction expansion coefficients $a$, $b$, $c$, $d$, $r$ \eqref{dual-interaction-log-expansion}. Recall that they are defined through the expansion of the dual interaction (in terms of its argument near zero):
\begin{equation}
        \ln\tilde{V}\left(\psi\right) = \ln\tilde{V}(0) - \frac{a}{2} \psi^{2}-\frac{b}{24}\psi^{4}-\frac{c}{6!} \psi^{6} -\frac{d}{8!} \psi^{8} - \frac{r}{10!}\psi^{10} +\ldots,
\end{equation}

 Here we assume that $\psi$ is a small parameter and scale out the coupling constant for brevity. Through algebraic simplification of Taylor Series expansion, one can deduce that:

\begin{equation}
\begin{aligned}
  a &=-\frac{\tilde{V}''(0)}{\tilde{V}(0)}>0, \qquad b=-\frac{\tilde{V}(0)\tilde{V}^{(4)}(0)-3\tilde{V}''(0)^{2}}{\tilde{V}(0)^{2}}, \\
    c &=-\frac{30\tilde{V}''(0)^{3}-15\tilde{V}(0)\tilde{V}^{(4)}(0)\tilde{V}''(0)+\tilde{V}(0)^{2}\tilde{V}^{(6)}(0)}{\tilde{V}(0)^{3}},
\end{aligned}
\end{equation}
and so on. Although the general expression can be derived using Faà di Bruno’s formula \cite{Stanley_combinatorics_2023}, it is not needed for our purposes. Note that all $\tilde{V}^{(2k+1)}(0)=0$ due to the parity of the original potential $V$. 

For power potentials, one can write down explicit formulas for derivatives:
\begin{equation}
\tilde{V}^{(2m)}(0)=2(-1)^{m}\int_{0}^{\infty}d\phi\, \phi^{2m}\exp\left[-\frac{1}{(2n)!}\phi^{2n}\right]=(-1)^m \frac{\Gamma \left(\frac{2 m+1}{2 n}\right) \Gamma (2 n+1)^{\frac{2 m+1}{2 n}}}{n},
\end{equation}

Therefore:
\begin{equation}
\label{equation:dual-action-expansion-coeffs}
\begin{aligned}
    a&=\frac{\Gamma \left(\frac{3}{2 n}\right) \Gamma (2 n+1)^{1/n}}{\Gamma \left(\frac{1}{2 n}\right)}>0,\qquad b=\frac{ \left(3 \Gamma \left(\frac{3}{2 n}\right)^2 - \Gamma \left(\frac{1}{2 n}\right) \Gamma \left(\frac{5}{2 n}\right)\right) \Gamma (2 n+1)^{2/n}}{\Gamma \left(\frac{1}{2 n}\right)^2}\geq0, \\
    c&=\frac{\left(30 \Gamma \left(\frac{3}{2 n}\right)^3-15 \Gamma \left(\frac{1}{2 n}\right) \Gamma \left(\frac{5}{2 n}\right) \Gamma \left(\frac{3}{2 n}\right)+\Gamma \left(\frac{1}{2 n}\right)^2 \Gamma \left(\frac{7}{2 n}\right)\right) \Gamma (2 n+1)^{3/n}}{\Gamma \left(\frac{1}{2 n}\right)^3}\geq0, \\
        d & = \frac{\Gamma (2 n+1)^{4/n}}{\Gamma \left(\frac{1}{2 n}\right)^4} 
    \left(7 \left(90 \Gamma \left(\frac{3}{2 n}\right)^4-60 \Gamma \left(\frac{1}{2 n}\right) \Gamma \left(\frac{5}{2 n}\right) \Gamma \left(\frac{3}{2 n}\right)^2 \right. \right. \\ & \left. \left. +4 \Gamma \left(\frac{1}{2 n}\right)^2 \Gamma \left(\frac{7}{2 n}\right) \Gamma \left(\frac{3}{2 n}\right)+5 \Gamma \left(\frac{1}{2 n}\right)^2 \Gamma \left(\frac{5}{2 n}\right)^2\right)-\Gamma \left(\frac{1}{2 n}\right)^3 \Gamma \left(\frac{9}{2 n}\right)\right) \geq 0
\end{aligned}
\end{equation}

and so on. For any given $n$, these coefficients can be calculated explicitly. We present these formulas here only for reference. For the $r$ coefficients we deliberately omit explicit formulas because of their size. The substitution shows that for $n=1$ the coefficients $b$, $c$ and all subsequent are set to zero, which is a corollary of the fact that the Fourier transform of a Gaussian is also Gaussian.

\begin{table}[h]
\caption{Numerical values of coefficients $a$, $b$, $c$, $d$, $r$ presented in \eqref{equation:dual-action-expansion-coeffs}.}
\label{table:dual-action-expansion-coeffs}
\centering
\begin{tabular}{|c|c|c|c|c|c|}
\hline
\textbf{n} & \textbf{a} & \textbf{b} & \textbf{c} & \textbf{d} & \textbf{r} \\  \hline
$2$          & $1.65580$    & $2.22504$    & $16.9726$    & $288.362$  & $8497.59$  \\ \hline
$3$          & $2.85398$    & $8.14521$    & $120.000$    & $3917.83$  & $221707.$ \\ \hline
$4$          & $4.45847$    & $21.4004$    & $513.351$    & $27230.2$  & $2.50252\cdot10^6$ \\ \hline
\end{tabular}
\end{table}

Numerical values of coefficients \eqref{equation:dual-action-expansion-coeffs} for specific potentials (e.g. $\phi^4$, $\phi^6$ and $\phi^8$) are summarized in Table~\ref{table:dual-action-expansion-coeffs}.

\section{Strong coupling background free energy per site calculation}
\label{appendix:strong-coupling-bg-free-en}

In this appendix, we derive the strong-coupling ``background'' free energy density in the thermodynamic limit \eqref{f0-strong-def} from the definition \eqref{F0-strong-def}.

To this end, we need the simplified expression for $\det \left( G + \frac{a}{g^2} \right)$. It is possible to write it down since we are interested in the limit $N\rightarrow \infty$. Moreover, we know the eigenvalues of $G$. We begin with the definition of the determinant and proceed with calculations, using \eqref{laplacian-eigensystem}, \eqref{green-func-raw}, and the Euler--Maclaurin summation formula, in the same way as during derivation of \eqref{weak-propagator}:

\begin{equation}
\begin{aligned}
        \frac{1}{N}\ln \det \left( G + \frac{a}{g^2} \right) & = \frac{1}{N} \sum_{k} \ln \left(\lambda_{k}^{-1} + \frac{a}{g^2} \right) \\ &= \frac{1}{(2\pi)^{d}} \left(\prod_{j=1}^d \int_{0}^{2 \pi} dq_{j}\right)\ln\frac{1 + \frac{a}{g^2}\left(4\alpha\sum_{j=1}^{d}\sin^{2}\left(\frac{q_{j}}{2}\right)+\gamma\right)}{4\alpha\sum_{j=1}^{d}\sin^{2}\left(\frac{q_{j}}{2}\right)+\gamma} + O\left(\frac{1}{N}\right),
\end{aligned}
\end{equation}

where we have intentionally placed the factor $1/N$ before the determinant to obtain an expression that remains finite in the limit $N\rightarrow\infty$. Here the sum over the Brillouin zone converges to the integral in thermodynamic limit. The expression under the logarithm is well-defined because we require all $\lambda_k$ to be positive.

Gathering all factors together, we obtain for ``background`` strong coupling free energy density $\tilde{f}_0$:

\begin{equation}
\label{f0-strong-appendix}
    \tilde{f}_0 =  \ln \frac{\sqrt{2\pi}}{\tilde{V}(0)} + \frac{1}{2 (2\pi)^{d}} \left(\prod_{j=1}^d \int_{0}^{2 \pi} dq_{j}\right)\ln\frac{g^2 + a\left(4\alpha\sum_{j=1}^{d}\sin^{2}\left(\frac{q_{j}}{2}\right)+\gamma\right)}{4\alpha\sum_{j=1}^{d}\sin^{2}\left(\frac{q_{j}}{2}\right)+\gamma}.
\end{equation}

It is important to note that this expression remains finite as $g\rightarrow 0$, and grows logarithmically, when $g\rightarrow \infty$, namely:

\begin{equation}
\label{f0-strong-asymptotic}
    \tilde{f}_0 =  \ln g  + \ln \frac{\sqrt{2\pi}}{\tilde{V}(0)} - \frac{1}{2 (2\pi)^{d}} \left(\prod_{j=1}^d \int_{0}^{2 \pi} dq_{j}\right)\ln \left[4\alpha\sum_{j=1}^{d}\sin^{2}\left(\frac{q_{j}}{2}\right)+\gamma \right] + O\left(\frac{1}{g^2} \right), \qquad g \rightarrow \infty.
\end{equation}

This asymptotic exhibits the expected concavity.

\section{Strong coupling Green's function calculation}
\label{appendix:strong-coupling-gf-calc}

In this appendix, we derive the Fourier Transform of dual theory Green's function \eqref{dual-theory-propagator-def}. A compact expression, in the spirit of \eqref{weak-propagator}, can be derived using \eqref{laplacian-eigensystem} and \eqref{green-func-raw}:
\begin{equation*}
\begin{split}
    \tilde{G}_{x,x^{\prime}} & =\left(G+\frac{a}{g^{2}}\right)^{-1}_{x,x'}=\left( \frac{L }{1+\frac{a}{g^{2}}L } \right)_{x,x'}=\frac{1}{N}\left(\prod_{i=1}^{d}\sum_{k_{i}=0}^{M-1}\right)\frac{4\alpha\sum_{j=1}^{d}\sin^{2}\left(\frac{p_{j}}{2}\right)+\gamma}{1+\frac{4 a\alpha}{g^{2}}\sum_{j=1}^{d}\sin^{2}\left(\frac{p_{j}}{2}\right)+\frac{a\gamma}{g^{2}}}e^{i\left(p,(x-x^{\prime})\right)}\\
 & =\frac{g^{2}}{a\left(2\pi\right)^{d}}\left(\prod_{i=1}^{d}\int_{0}^{2\pi}dq_{i}\right)\frac{4 \alpha \sum_{j=1}^{d}\sin^{2}\left(\frac{q_{j}}{2}\right)+\gamma}{4\alpha \sum_{j=1}^{d}\sin^{2}\left(\frac{q_{j}}{2}\right)+\left(\frac{g^{2}}{a}+\gamma\right)}e^{i\left(q,x-x^{\prime}\right)}
\end{split}
\end{equation*}
The last equality is the application of the spectral theorem and substitution of explicit value of lattice Laplacian (Appendix~\ref{appendix:laplacian}). In our derivations, we follow the conventions of standard lattice field theory textbooks (e.g., \cite{Montvay1997-be}).

We will use the same shorthand notations for $\tilde{G}$ matrix elements as for $G$, namely, $\tilde{G}_x = \tilde{G}_{x,0} = \tilde{G}(x)$ and the same for the Fourier transform:

\begin{equation}
\label{dual-theory-propagator-appendix}
     \tilde{G}(q) =\frac{g^2}{a}\frac{4\alpha \sum_{j=1}^{d}\sin^{2}\left(\frac{q_{j}}{2}\right)+\gamma}{4\alpha\sum_{j=1}^{d}\sin^{2}\left(\frac{q_{j}}{2}\right)+\left(\frac{g^{2}}{a}+\gamma\right)}.
\end{equation}

The physical interpretation of this result is not immediately transparent. Furthermore, note that $\tilde{G}$ vanishes as $g\rightarrow 0$, and approaches a constant (depending on momentum) as $g \rightarrow \infty$. This behaviour renders all terms regular in the limit $g\to0$, which substantially improves subsequent perturbation finite-order behavior.

\section{Perturbative calculation of Free energy}
\label{appendix:pert-free-energy-calc}

In this appendix, we detail the calculation of the free energy per site \eqref{free-energy-per-site-def}. Using the definition \eqref{free-energy-def}, we perform a perturbative expansion in powers of $g$ using standard Feynman diagram techniques, considering only connected ``vacuum bubbles'' (vacuum diagrams). We will provide the two different perturbative expansions of free energy. The first one is a common weak coupling expansion with standard Feynman technique for action \eqref{lattice-action}. The second one is a strong coupling expansion for the initial theory, obtained as a weak coupling expansion of a dual theory \eqref{dual-action-shortened-expanded}, which involves a different set of Feynman diagrams due to the distinct structures of the kinetic and interaction terms in the action. The Feynman diagram technique is common in lattice field theory \cite{Montvay1997-be, Smit_2002-lattice}.

\subsection{Weak coupling case}

\begin{figure}
    \centering
    \includegraphics[width=0.9\linewidth]{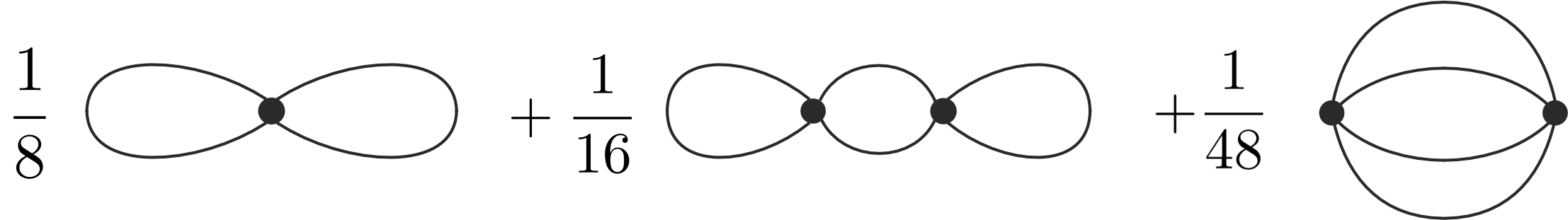}
    \caption{Feynman Graph for the initial theory \eqref{lattice-action} in momentum space contributing to the free energy per site \eqref{free-energy-per-site-def} up to order $g^8$, written down in the expression \eqref{free-energy-weak}. Black single lines correspond to the initial theory propagator \eqref{weak-propagator-ft}. Black round vertices correspond to interaction $V(\phi)=\frac{g^4}{4!}\phi^4$, resulting in the factor $- (2 \pi)^d g^4 \cdot \delta_{\sum p_i, 0}$, where the last Kronecker symbol reflects momentum conservation in each vertex.}
    \label{fig:f-weak-diagrams}
\end{figure}

Starting from the partition function \eqref{PartFunc1}, we can provide its log Feynman perturbation expansion. Summing the contributions of all diagrams shown in Figure~\ref{fig:f-weak-diagrams}, we obtain the weak-coupling perturbative expression for the free energy per site \eqref{free-energy-per-site-def}:

\begin{equation}
\label{free-energy-weak}
\begin{split}
f & = \frac{g^{4}}{8}\left(G_{0}\right)^{2}-\frac{1}{16(2\pi)^{d}}g^{8} \left(G_0\right)^{2}\left(\prod_{i=1}^{d}\int_{0}^{2\pi}dq_{i}\right)G(q)^{2}\\
 & -\frac{g^{8}}{48(2\pi)^{3d}}\left(\prod_{n=1}^{3}\left(\prod_{i=1}^{d}\int_{0}^{2\pi}dq_{i}^{n}\right)G(q^{n})\right)G\left(q^{1}+q^{2}+q^{3}\right)+O\left(g^{12}\right),
 \end{split}
\end{equation}

following the notations introduced in \eqref{GF-shorthand-notations} and \eqref{weak-propagator-ft}. Note that the resulting expression for $f$ remains finite in the limit $N\rightarrow\infty$. 

\subsection{Strong coupling case}

\begin{figure}
    \centering
    \includegraphics[width=0.9\linewidth]{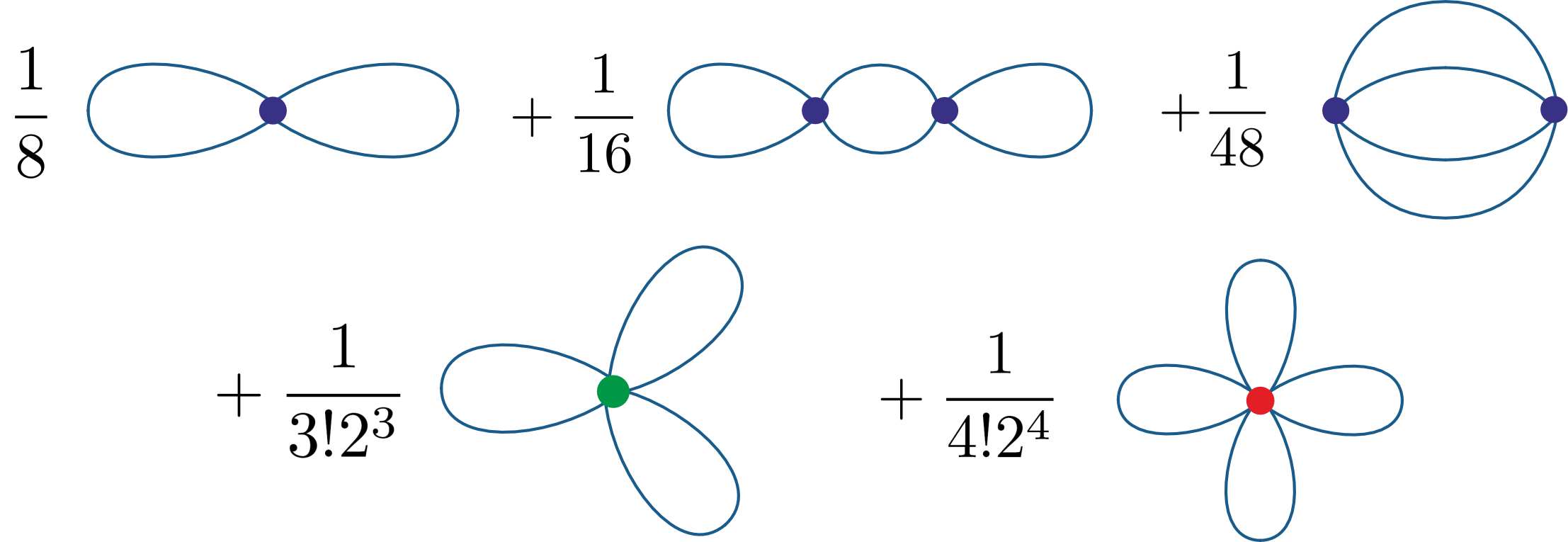}
    \caption{Feynman Graphs for the dual theory \eqref{dual-action-shortened-expanded} in momentum space up to order $\frac{1}{g^8}$ contributing to the free energy per site \eqref{free-energy-per-site-def} from \eqref{free-energy-strong}. Blue single lines correspond to the dual theory \eqref{dual-action-expanded} propagator \eqref{dual-theory-propagator}. Blue round vertices correspond to interaction $V_4^*(\psi)=\frac{b}{4!g^4}\psi^4$. Similarly, green round vertices reflect the term  $V_6^*(\psi)=\frac{c}{6!g^6}\psi^6$ and the red ones - the term $V_8^*(\psi)=\frac{d}{8!g^8}\psi^8$. The coefficients $b$, $c$ and $d$ are given by \eqref{equation:dual-action-expansion-coeffs}.}
    \label{fig:f-strong-diagrams}
\end{figure}

In this section, we develop a perturbative expansion in the strong-coupling regime, i.e. in powers of $1/g$, which is complementary to the conventional weak-coupling expansion. Our aim is to compute the first few non-trivial terms of the expansion—specifically, those up to order $(1/g)^8$.

To this end, we begin with the dual representation of the partition function given in Eq.~\eqref{dual-partition-general-final}. Throughout the calculation, we retain only contributions of order $(1/g)^4$, $(1/g)^6$, and $(1/g)^8$.

Strictly speaking, a systematic expansion would require decomposing the propagator itself in powers of $1/g$. To avoid excessively cumbersome intermediate expressions, we adopt a resummed approach: all quadratic (Gaussian) contributions in the action are kept intact, effectively summing them to all orders in $1/g$. In the language of perturbation theory, this amounts to working with a dressed propagator that already incorporates the full $1/g$ dependence at the quadratic level. This approach offers an additional advantage: all resulting expressions remain regular in the limit $g \to 0$. This is due to the explicit $g$-dependence of the propagator, which suppresses its magnitude as the coupling becomes small.

With this prescription in place, we now proceed to the explicit computation of the higher-order corrections.

Exactly as in the case of weak coupling limit, we are interested only in free energy $\mathcal{F}$, which is given by the connected diagrams only. Unlike the perturbation series for the initial theory, the dual action has the interaction consisting of infinitely many powers of dual field $\psi$. While straightforward, the computation becomes algebraically cumbersome at higher orders. The necessary Feynman diagrams are presented in Figure~\ref{fig:f-strong-diagrams}. So let us write down the first few terms for $f$ explicitly without diving into the computations: 

\begin{equation}
\label{free-energy-strong}
    \begin{split}
 f & =\tilde{f}_0+\frac{b}{8 g^{4}}(\tilde{G}_0)^{2}+\frac{c}{2^{3}3! g^{6}}(\tilde{G}_0)^{3}-\frac{b^{2}}{2^{4}(2\pi)^{d} g^{8}}\left((\tilde{G}_0)^{2}\left(\prod_{i=1}^{d}\int_{0}^{2\pi}dq_{i}\right)\tilde{G}(q)^{2}\right)\\
 & + \frac{d}{2^4 4! g^8} \left(\tilde{G}_0\right)^4 -\frac{b^{2}}{48 (2\pi)^{3d} g^{8}}\left(\prod_{n=1}^{3}\left(\prod_{i=1}^{d}\int_{0}^{2\pi}dq_{i}^{n}\right)\tilde{G}\left(q^{n}\right)\right)\tilde{G}\left(q^{1}+q^{2}+q^{3}\right) + O\left(1/g^{10}\right),
    \end{split}
\end{equation}

where $\tilde{f}_0$ is given by \eqref{f0-strong}. Let us note that the coefficients $a$, $b$, $c$ and $d$ are the dual action coefficients and can be found in Appendix~\ref{appendix:dual-action-coeffs} with numerical values summarized in Table~\ref{table:dual-action-expansion-coeffs} (the case under consideration corresponds to $n=2$).

\section{Two-point function}
\label{appendix:two-point}

One can evaluate the two-point function for the initial and the dual theory in momentum space in terms of \eqref{DFT-inverse} and \eqref{two-point-momentum-def} with the help of Feynman Perturbation Theory. The calculations are the same as in the case of free energy density, except for the fact that now all the diagrams will have external vertices (in coordinate picture) or external legs (in momentum space). It is more convenient to work in momentum space, as the free theory ($g=0$) result can be expressed in terms of elementary functions.
\begin{figure}
    \centering
    \includegraphics[width=\linewidth]{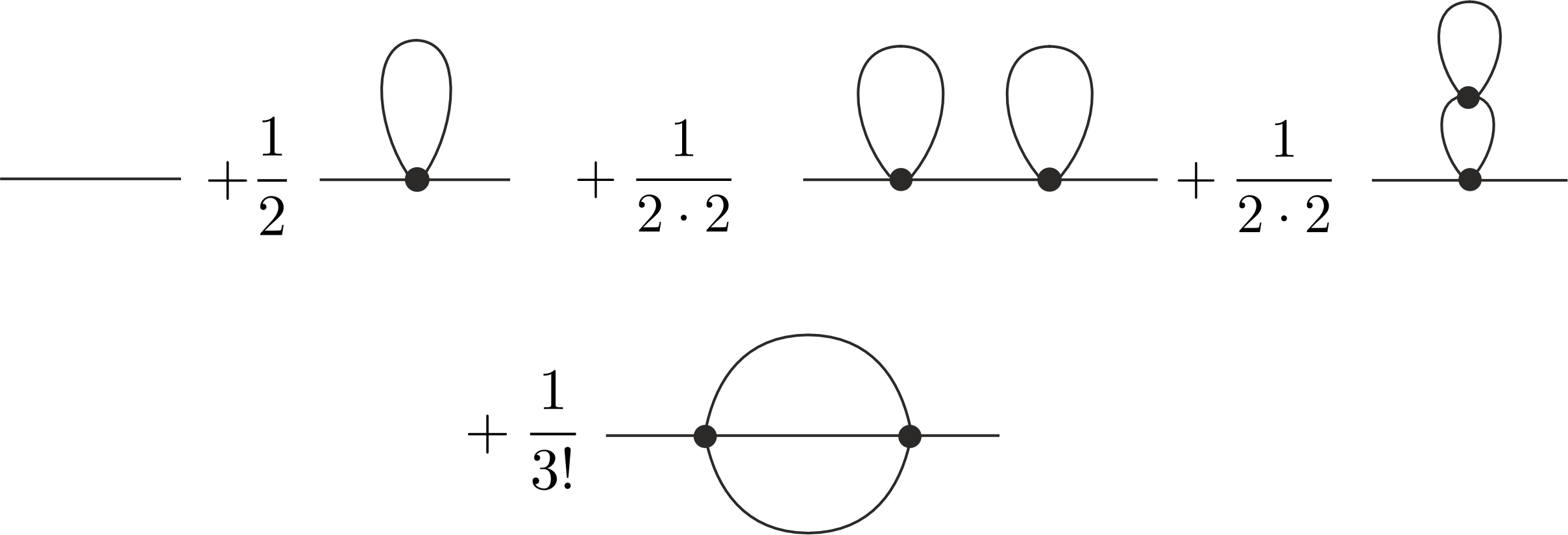}
    \caption{Feynman Graphs for the two-point function in momentum space \eqref{two-point-momentum-def} in the initial theory \eqref{lattice-action} up to order $g^8$, written down in \eqref{two-point-weak}. The diagram technique is the same as in Figure~\ref{fig:f-weak-diagrams}.}
    \label{fig:feynman-graphs-two-point-weak}
\end{figure}

For the initial theory, one should sum all the diagrams from Figure~\ref{fig:feynman-graphs-two-point-weak}. It is the standard computation, so let us write down the final answer for two-point function \eqref{two-point-momentum-def} without going into the details:

\begin{equation}
\label{two-point-weak}
\begin{split}
    \frac{\left\langle \phi(p)\phi(-p)\right\rangle}{G(p)^2} & = \frac{1}{G(p)}-\frac{g^{4}}{2} G_0 + \frac{g^8}{2^2} \left(G_0\right)^2 G(p)  + \frac{g^8}{2^2 (2\pi)^{d}} G_0 \left(\prod_{i=1}^{d}\int_{0}^{2\pi}dq_{i}\right) G(q)^2 \\ &+\frac{g^{8}}{3!(2\pi)^{2d}} \left(\prod_{i=1}^{d}\int_{0}^{2\pi}dq_{i}\right)\left(\prod_{i=1}^{d}\int_{0}^{2\pi}dk_{i}\right) G(k) G(q) G(q+k + p) + O(g^{12}), \qquad g \rightarrow 0
\end{split}
\end{equation}

\begin{figure}
    \centering
    \includegraphics[width=\linewidth]{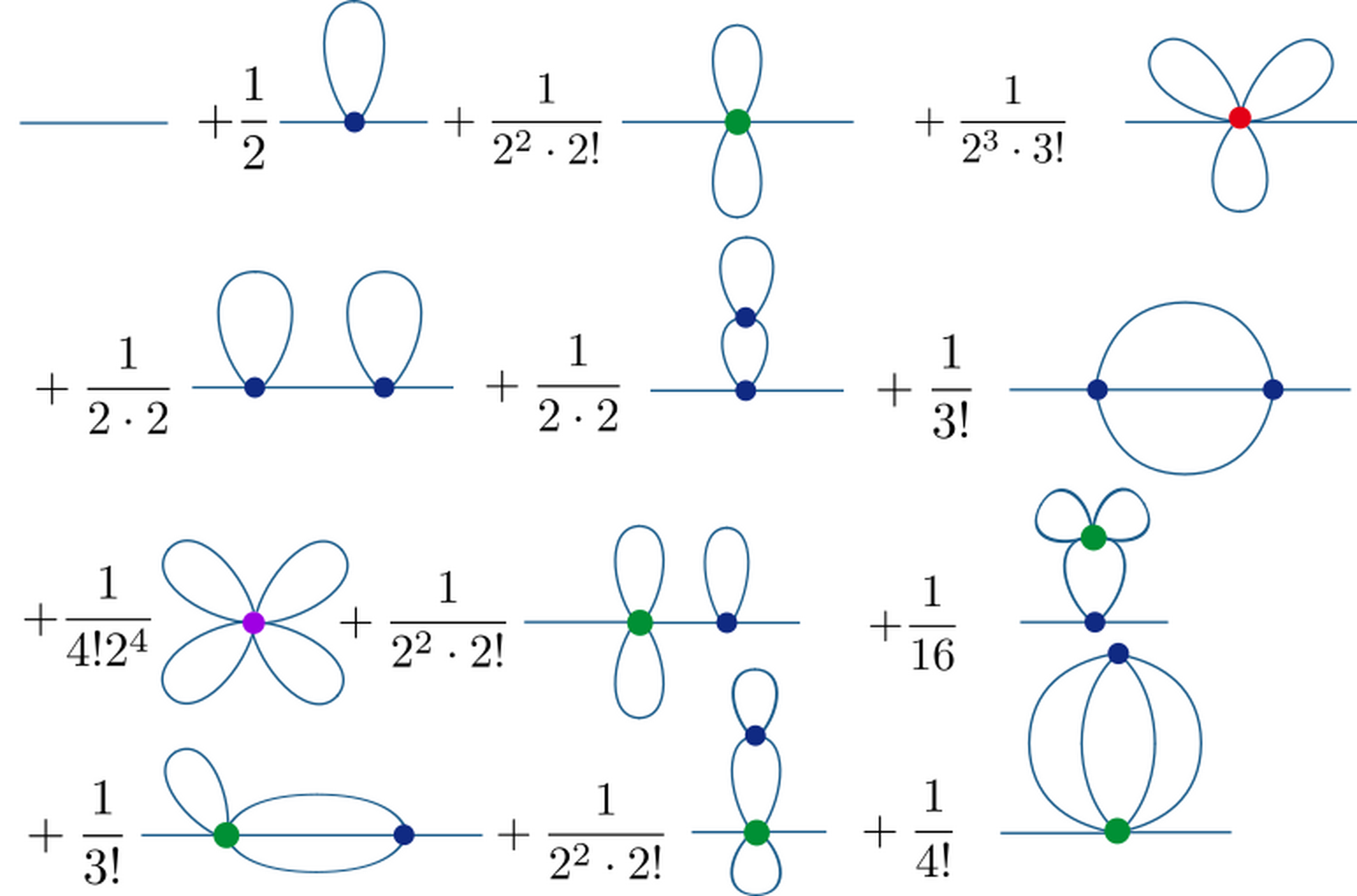}
    \caption{Feynman Graphs for the two-point function in momentum space \eqref{two-point-momentum-def} in the dual theory \eqref{dual-action-shortened-expanded} up to order $\frac{1}{g^{10}}$, written down in \eqref{two-point-strong-appendix}. Blue single lines correspond to the dual theory \eqref{dual-action-expanded} propagator \eqref{dual-theory-propagator}. Blue round vertices correspond to interaction $V_4^*(\psi)=\frac{b}{4!g^4}\psi^4$. Green round vertices reflect the term  $V_6^*(\psi)=\frac{c}{6!g^6}\psi^6$, the red ones - the term $V_8^*(\psi)=\frac{d}{8!g^8}\psi^8$ and the violet - the term  $V_{10}^*(\psi)=\frac{r}{10!g^{10}}\psi^{10}$. For the mixed $bc$ topologies with one external leg attached to each inequivalent vertex, the displayed factor includes the equal contribution obtained by interchanging the two external legs. The coefficients $b$, $c$, $d$ and $r$ are given by \eqref{equation:dual-action-expansion-coeffs}. }
    \label{fig:feynman-graphs-two-point-strong}
\end{figure}

For the dual theory, the correlation function \eqref{two-point-momentum-def} can be computed by accounting for all contributions shown in Figure~\ref{fig:feynman-graphs-two-point-strong}. Let us introduce the notations for momentum integrals as:

\begin{equation}
\label{integrals-two-point-appendix}
\begin{split}    
    \tilde{I}_2 &= \frac{1}{(2 \pi)^d}\left(\prod_{i=1}^{d}\int_{0}^{2\pi}dq_{i}\right) \tilde{G}(q)^2, \\
    \tilde{I}_3(p) &= \frac{1}{(2\pi)^{2d}} \left(\prod_{i=1}^{d}\int_{0}^{2\pi}dq_{i}\right)\left(\prod_{i=1}^{d}\int_{0}^{2\pi}dk_{i}\right) \tilde{G}(k) \tilde{G}(q) \tilde{G}(q+k + p), \\
    \tilde{I}_4 &= \frac{1}{(2\pi)^{3d}} \left(\prod_{n=1}^{3}\left(\prod_{i=1}^{d}\int_{0}^{2\pi}dq_{i}^{n}\right)\tilde{G}\left(q^{n}\right)\right)\tilde{G}\left(q^{1}+q^{2}+q^{3}\right).
\end{split}
\end{equation}

Then the final formula reads:

\begin{equation}
\label{two-point-strong-appendix}
\begin{split}
\frac{\left\langle \psi(p)\psi(-p)\right\rangle^* }{\tilde{G}(p)^2} & =\frac{1}{\tilde{G}(p)}-\frac{b}{2 g^4}\tilde{G}_0 - \frac{c}{2^3 g^6} \left(\tilde{G}_0\right)^2 - \frac{d}{3! 2^3 g^{8}}\left( \tilde{G}_0 \right)^3 + \frac{b^2}{2^2 g^8} \tilde{G}_0 \tilde{I}_2 +\frac{b^2}{2^2 g^8} \left(\tilde{G}_0\right)^2 \tilde{G}(p) \\ & + \frac{b^2}{3! g^8 } \tilde{I}_3 (p) -\frac{r}{4!2^4g^{10}}\left(\tilde{G}_0\right)^4 + \frac{bc}{8 g^{10}} \left(\tilde{G}_0\right)^3 \tilde{G}(p) + \left( \frac{1}{8} + \frac{1}{16} \right) \frac{bc}{g^{10}} \left(\tilde{G}_0\right)^2\ \tilde{I}_2
 \\& +\frac{bc}{6g^{10}} \tilde{G}_0 \tilde{I}_3(p) +\frac{bc}{24 g^{10}} \tilde{I_4} + O\left(\frac{1}{g^{12}}\right) , \qquad g \rightarrow \infty.
\end{split}
\end{equation}

Although the amputated quantity above scales as $g^{-2}$, multiplication by $\widetilde G(p)^2$ yields $\left\langle \psi(p)\psi(-p)\right\rangle^*=O(g^2)$; hence the full dual two-point function remains regular as $g\to0$ exactly as in the case of free energy per site $f$. Now, using the relation \eqref{correlators-link}, we are able to obtain the strong coupling expansion for the two-point correlation function $\langle \phi(p)\phi(-p) \rangle $.

It is instructive to explicitly write down the strong-coupling Green's function derived from \eqref{two-point-strong} and \eqref{correlators-link}. Substituting, we obtain to the first few orders:

\begin{equation}
     \langle \phi(p)\phi(-p)\rangle = \frac{1}{4\alpha\sum_{j=1}^{d}\sin^{2}\left(\frac{p_{j}}{2}\right)  +\frac{g^2}{a} + \gamma}+ \frac{b\tilde{G}_0}{2a^2} \frac{1}{ \left( \gamma + 4 \alpha  \sum_{j=1}^{d}\sin^{2}\left(\frac{p_{j}}{2}\right) + \frac{g^2}{a}\right)^2} +  O\left(\frac{1}{g^{6}}\right), \qquad g \rightarrow \infty.
\end{equation}

Remarkably, the non-essential factors cancel out, resulting in the structure of the standard $\phi^4$ perturbation theory with a Laplacian-type propagator, similar to \eqref{green-func-raw}, but with the effective parameters. For instance, now it is reasonable to call the combination:

\begin{equation}
    \tilde{m}^2 = \frac{g^2}{a} + \gamma,
\end{equation}

an effective squared mass, compared to the bare mass $m^2 = \gamma$.

A potential subtlety arises at the point $\tilde{m}^2=0$, where the propagator exhibits a singularity. However, we refrain from premature conclusions, as the expansion was derived for the strong-coupling regime ($g\rightarrow\infty$); moreover, the condition $\tilde{m}^2=0$ implies $\gamma = -g^2 / a$ makes the kinetic operator of the initial theory not positive and therefore violates the positivity assumptions under which the duality was derived. The analytical continuation of the strong-coupling expansion to the physical regime requires further investigation. While the poles of the Green's function obtained within the strong-coupling expansion, when augmented by perturbative corrections, hint at a viable approach for spectral analysis, a rigorous justification of this continuation remains a significant task for the future research.

\section{Numerical methods used for simulations}
\label{appendix:num-methods}

We now test the obtained duality through the numerical simulation results. The simulation procedure itself is standard and is not the focus of the present work. We do not describe the standard HMC methodology in detail here. Instead, we summarize
only the parameters and analysis choices relevant for reproducibility and for
the comparison with the analytical results. Implementation details are
available in the accompanying code repository.
The code used to generate the numerical results and analytical comparison curves, together with the simulation configurations and reference data used in this work, is available at \cite{LatticePhi4_2023}.

We are going to compare numerical and analytical (both strong and weak expansions) for free-energy curves as functions of $g$ and two-point function in momentum space.

\subsection{Numerical calculations of free energy and correlators}
\label{numerical-sum-sect}

The numerical simulations are performed using the Hamiltonian Monte Carlo (HMC) method \cite{HMC_START_DUANE, Modern_Perspectives_in_Lattice_QCD,neal2012mcmc}. This approach was first introduced in \cite{HMC_START_DUANE}. We prefer this approach over local Metropolis algorithm, as it generally ensures faster convergence and is relatively straightforward to implement.

Hamiltonian Monte Carlo method is an improvement of the Metropolis-Hastings algorithm. The main difference is that one includes additional integrations over some momenta and then considers the obtained function in the exponent as a Hamiltonian. HMC augments the field variables with auxiliary conjugate momenta and proposes global updates by approximately integrating Hamilton's equations in the enlarged phase space. A Metropolis accept/reject step removes the bias caused by the finite integration step. This approach reduces autocorrelation times, making calculations more efficient \cite{PhysRevD.96.114505_lattice_hmc_simulations_1,Pawlowski_2020_lattice_hmc_simulations_2}.

The used implementation of the described algorithm can be found in \cite{LatticePhi4_2023}. This repository is based on a pedagogical example from \cite{LatticePhi4_root} (being the illustration to works \cite{PhysRevD.96.114505_lattice_hmc_simulations_1,Pawlowski_2020_lattice_hmc_simulations_2}). All the presented simulations were performed on a periodic hypercubic lattice of unit spacing with $M=32$ sites in each direction, i.e. $N=32^d$ sites in total. More precise information about simulation parameters can be found in Tables \ref{tab:hmc_parameters_free_energy} and \ref{tab:hmc_parameters_two_point}. 

\begin{table}[t]
    \centering
    \caption{HMC simulation parameters used for the free-energy per site calculations.}
    \label{tab:hmc_parameters_free_energy}
    \begin{tabular}{|c|c|c|c|c|c|}
        \toprule
        Dimension $d$
        & Warm-up steps
        & Production steps
        & Sampling interval
        & Recorded samples
        & Base Leapfrog steps \\
        \midrule
        2 & $1\,000$  & $10\,000$ & 10 & $1\,000$ & 100 \\ 
        3 & $10\,000$ & $20\,000$ & 10 & $2\,000$ & 100 \\
        \bottomrule
    \end{tabular}
\end{table}

\begin{table}[t]
    \centering
    \caption{HMC simulation parameters used for the two-point-function calculations.}
    \label{tab:hmc_parameters_two_point}
    \begin{tabular}{|c|c|c|c|c|c|}
        \toprule
        Dimension $d$
        & Warm-up steps
        & Production steps
        & Sampling interval
        & Recorded samples
        & Base Leapfrog steps \\
        \midrule
        2 & $5\,000$  & $60\,000$ & 10 & $6\,000$ & 20 \\ 
        3 & $10\,000$ & $70\,000$ & 10 & $7\,000$ & 30 \\
        \bottomrule
    \end{tabular}
\end{table}

Standard Monte Carlo methods do not allow direct computation of the partition function; however, they are efficient for calculating correlation functions. This is not a significant obstacle, as one can observe that:

\begin{equation}
    \frac{\partial f}{\partial (g^4)} = \frac{1}{4!} \left< \phi^4 (x)\right> 
\end{equation}

for any fixed site $x$ (due to the translational invariance of the lattice under consideration with periodic boundary conditions). So, we can deduce that:

\begin{equation}
\label{appendix-free-energy-integration}
    f(g) = \frac{1}{4!} \int\limits_0^{g^4}  \left< \phi^4 (x)\right>_{g^4=\mu} d\mu,
\end{equation}

using $f(g=0) = 0$, where correlators in integrand are considered as functions of coupling constant and the integration variable $\mu=g^4$ has been introduced for brevity. The numerical integration is performed using SciPy (v. 1.13.0) Python 3 package \cite{SciPy} implementation of quadrature integration after cubic spline interpolation.

Furthermore, the expectation value of an operator can be estimated from the distribution of numerically generated field configurations $\Phi$ as:

\begin{equation}
\left< \phi^4 (x)\right> \approx \frac{1}{|\Phi|}\sum\limits_{\phi \in \Phi} \frac{1}{N} \sum\limits_{x} \phi(x)^4,
\end{equation}

where we also used the averaging along the lattice sites to increase the precision.

Similarly, for the two-point correlation function, one can write:

\begin{equation}
    \left< \phi (x_1) \phi(x_2)\right> \approx \frac{1}{|\Phi|}\sum\limits_{\phi \in \Phi} \phi (x_1)\phi(x_2).
\end{equation}

To obtain the correlators in momentum space $\langle \phi(p)\phi(-p) \rangle $ as a function of admissible $p$, one can take the Discrete Fourier Transform \eqref{DFT} for a grid of $p$.

\subsection{Error estimation}
\label{appendix-num-error-estimation}

The error for correlators obtained with the provided HMC simulation was estimated using the blocked jackknife over consecutive retained configurations \cite{Modern_Perspectives_in_Lattice_QCD}. Blocking is used to account for autocorrelations along the Markov chain. For the two-point function, the
jackknife procedure is applied independently to each lattice momentum.

\begin{figure}

    \centering
    \includegraphics[width=8cm,height=5.5cm]{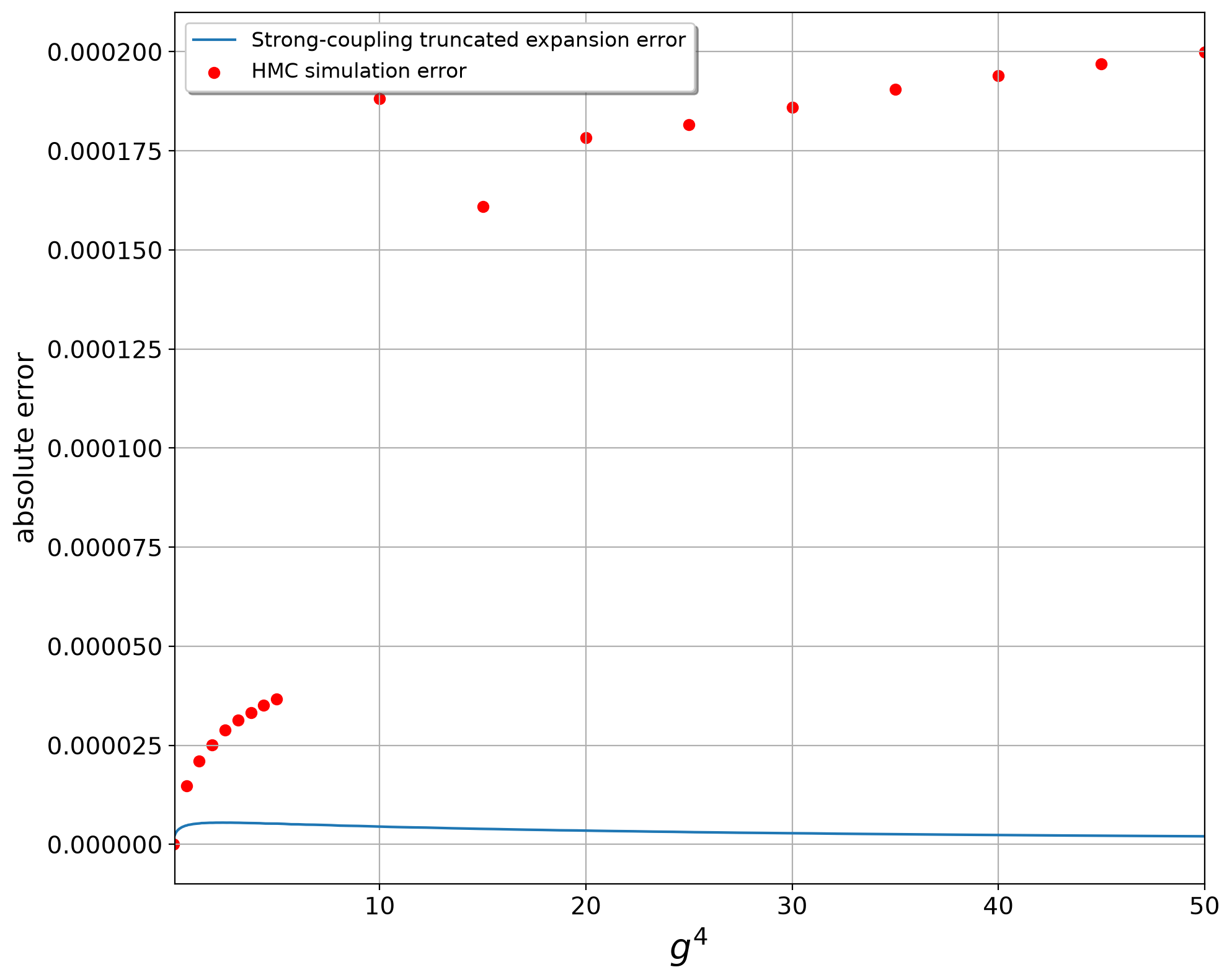}
    \hfill
    \includegraphics[width=8cm,height=5.5cm]{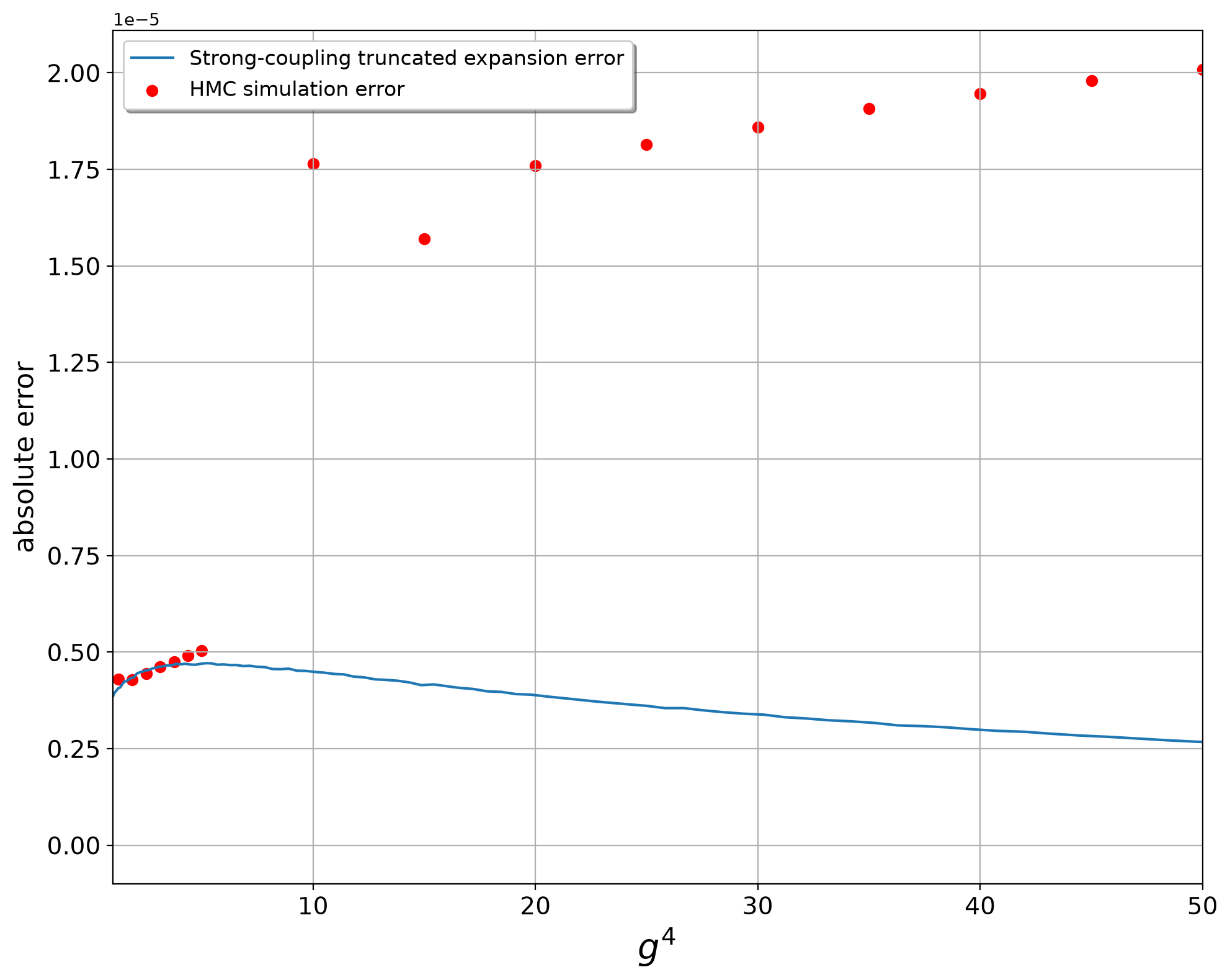}
    \caption{Plots of errors for analytical (strong coupling) and computational results for the free energy per site \eqref{free-energy-per-site-def}. Red points represent the error of free energy per site caused by HMC simulations error. Blue curve is the numerical integration error of strong coupling analytical expansions. Left panel is for $d=2$ and the right one is for $d=3$.}
    \label{fig:err-f(g)-comparison}
\end{figure}

The free energy per site is obtained by thermodynamic integration~\eqref{appendix-free-energy-integration}. The measured values of $\langle\phi^4\rangle$ are interpolated as described
above. Since both interpolation and integration are linear in the measured
values, the statistical uncertainty of the resulting free energy can be
propagated directly. Writing
\begin{equation}
  f(g_i) =
\frac{1}{24}
\sum_j w_j(g_i)
\left\langle\phi^4\right\rangle_j ,  
\end{equation}

where $w_j(g_i)$ are the integration weights associated with the interpolation
scheme, we estimate
\begin{equation}
\sigma_f^2(g_i)
=
\frac{1}{4!^2}
\sum_j
w_j^2(g_i)\,
\sigma_j^2 ,    
\end{equation}

assuming statistically independent HMC runs at different coupling values.

\begin{figure}
    \centering
    \includegraphics[width=8cm,height=5.5cm]{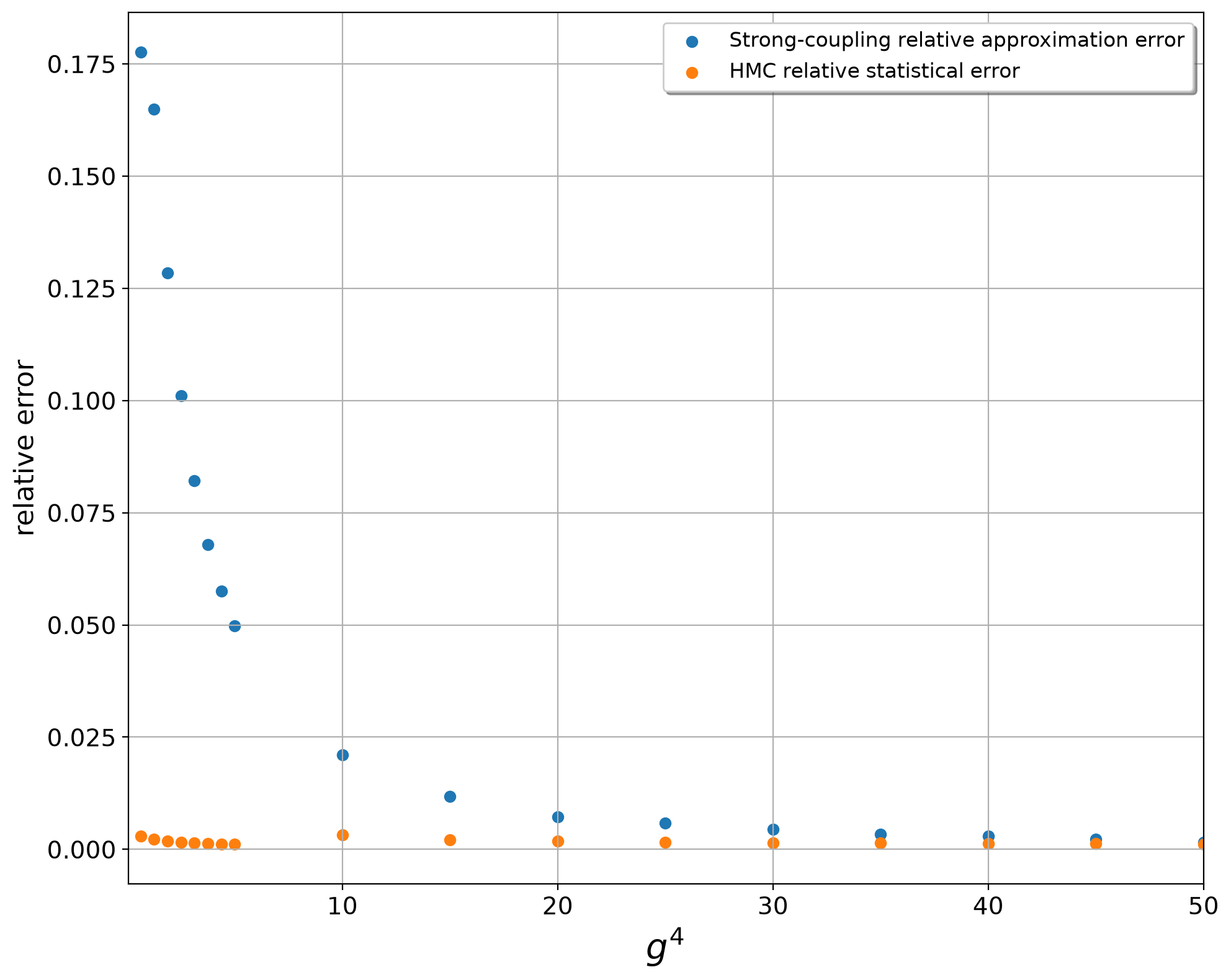}
    \hfill
    \includegraphics[width=8cm,height=5.5cm]{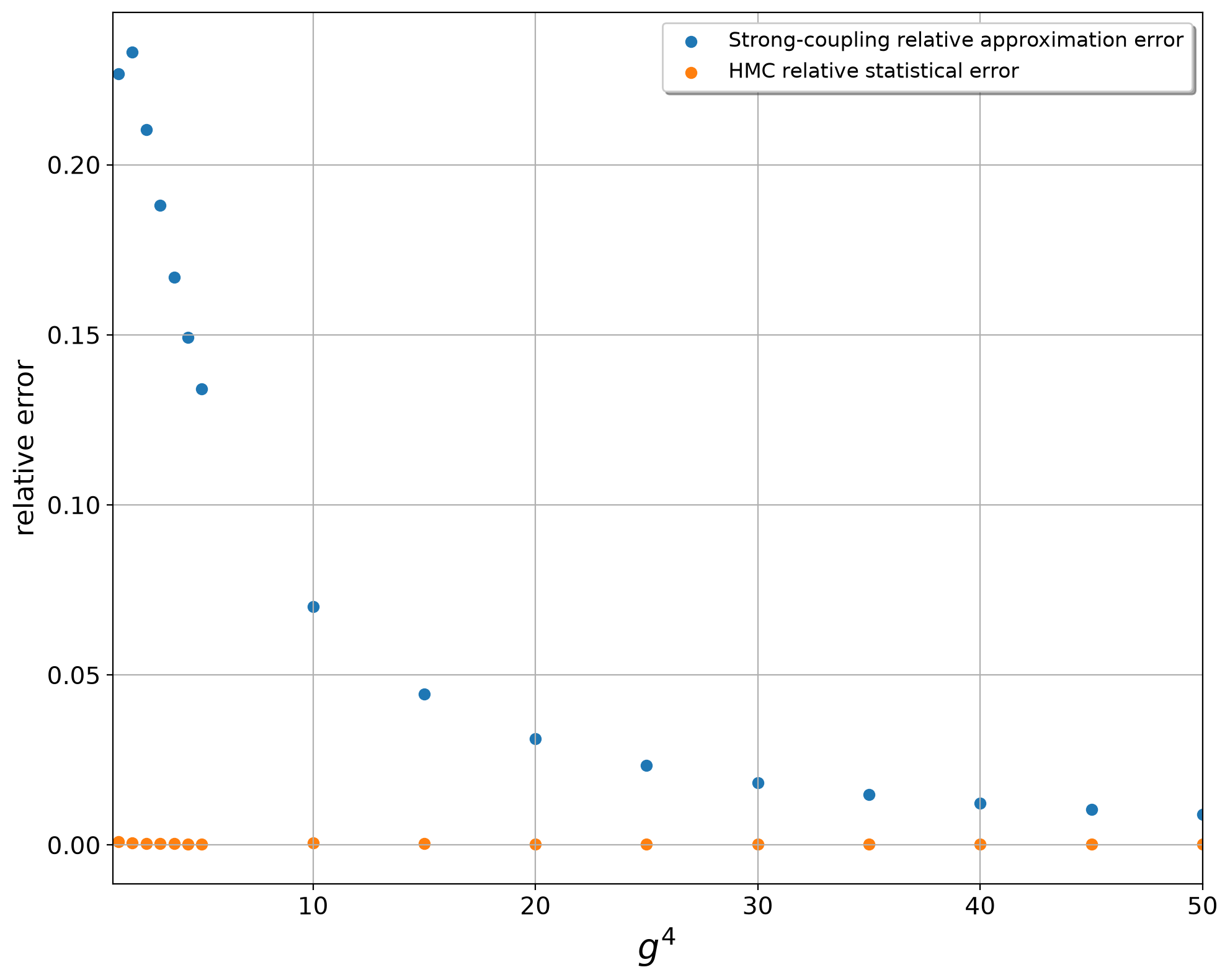}
    \caption{Relative HMC simulation error
$\epsilon_{\mathrm{rel,HMC}}(g)=\epsilon_{\mathrm{HMC}}(g)/f_{\mathrm{HMC}}(g)$
and relative error of the strong-coupling expansion truncated at order $g^{-8}$,
$\epsilon_{\mathrm{rel,approx}}(g)=|f_{\mathrm{HMC}}(g)-f_{\mathrm{approx}}(g)|/f_{\mathrm{HMC}}(g)$,
for the free energy per site \eqref{free-energy-per-site-def}.
Here $f_{\mathrm{HMC}}(g)$ is the HMC estimate of the free energy per site,
$\epsilon_{\mathrm{HMC}}(g)$ is its absolute error,
and $f_{\mathrm{approx}}(g)$ is the strong-coupling approximation
\eqref{free-energy-strong}.
The left panel corresponds to $d=2$ and the right panel to $d=3$.}
    \label{fig:f(g)-relative-errors}
\end{figure}

The integrals in the analytical expressions \eqref{free-energy-strong}, \eqref{free-energy-weak} and others were computed numerically using quadratures (for number of integration dimensions 1 or 2) or Monte Carlo integration (for 3 and more integration dimensions). The errors of Monte Carlo calculations were also estimated using the jackknife approach, the error of quadrature integration was given by the Python 3 SciPy package (v. 1.13.0) \cite{SciPy}. The Python 3 library NumPy \cite{NumPy} has also been used for array manipulation. The absolute errors of the calculations performed in region $g \geq 1$ for $\gamma = 1$  and $\alpha = 1$ are presented in Figure~\ref{fig:err-f(g)-comparison}. For the calculation of the Feynman integrals with Monte Carlo approach, we used $5\cdot 10^4$ points for $d=2$ and $d=3$. The graphs demonstrate that the error in the calculations performed remains sufficiently small. Additionally, it does not introduce any ambiguity to the check that truncated series approaches the numerical result over the tested range.

\begin{figure}
    \centering
    \includegraphics[width=8cm,height=5.5cm]{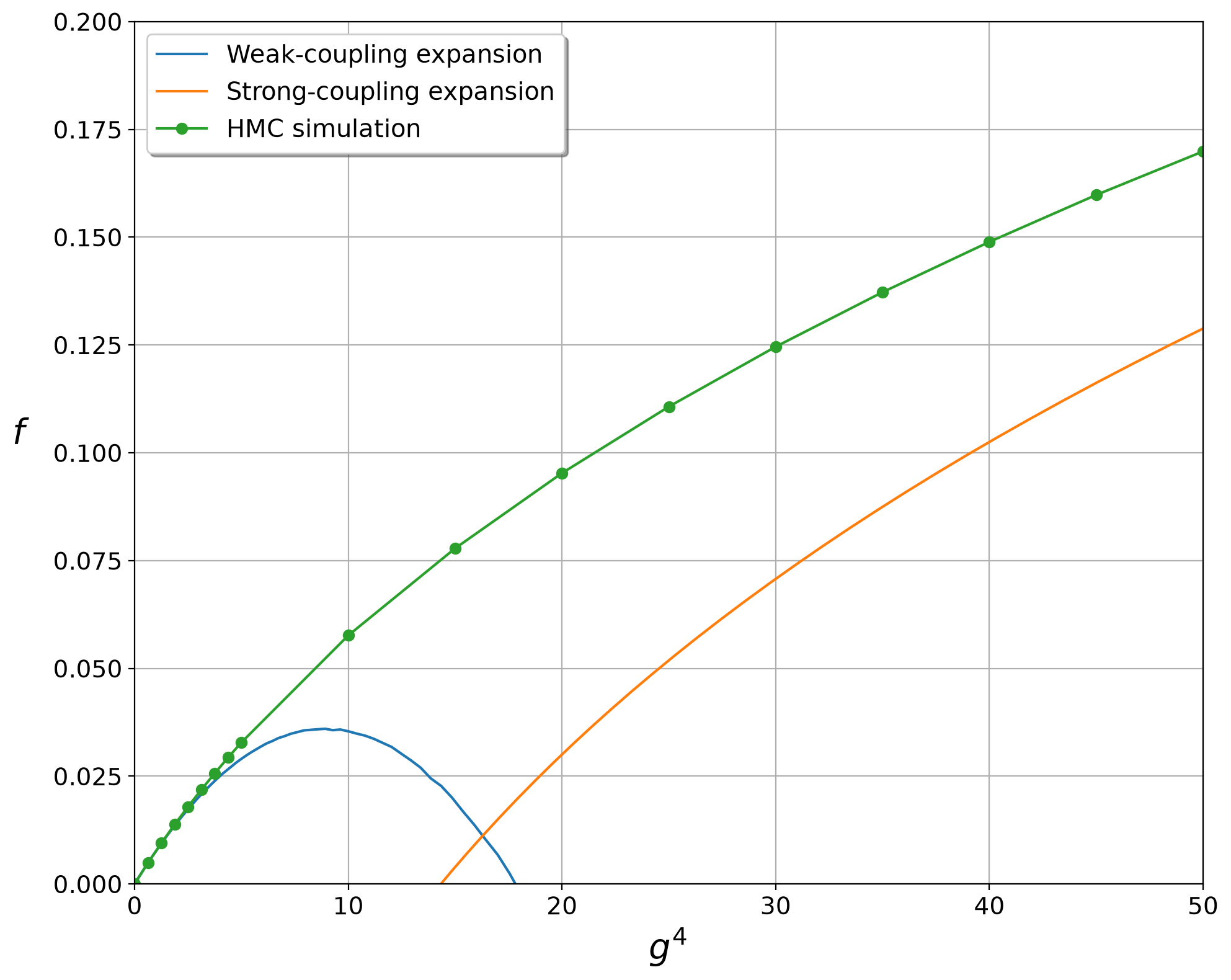}
    \hfill
    \includegraphics[width=8cm,height=5.5cm]{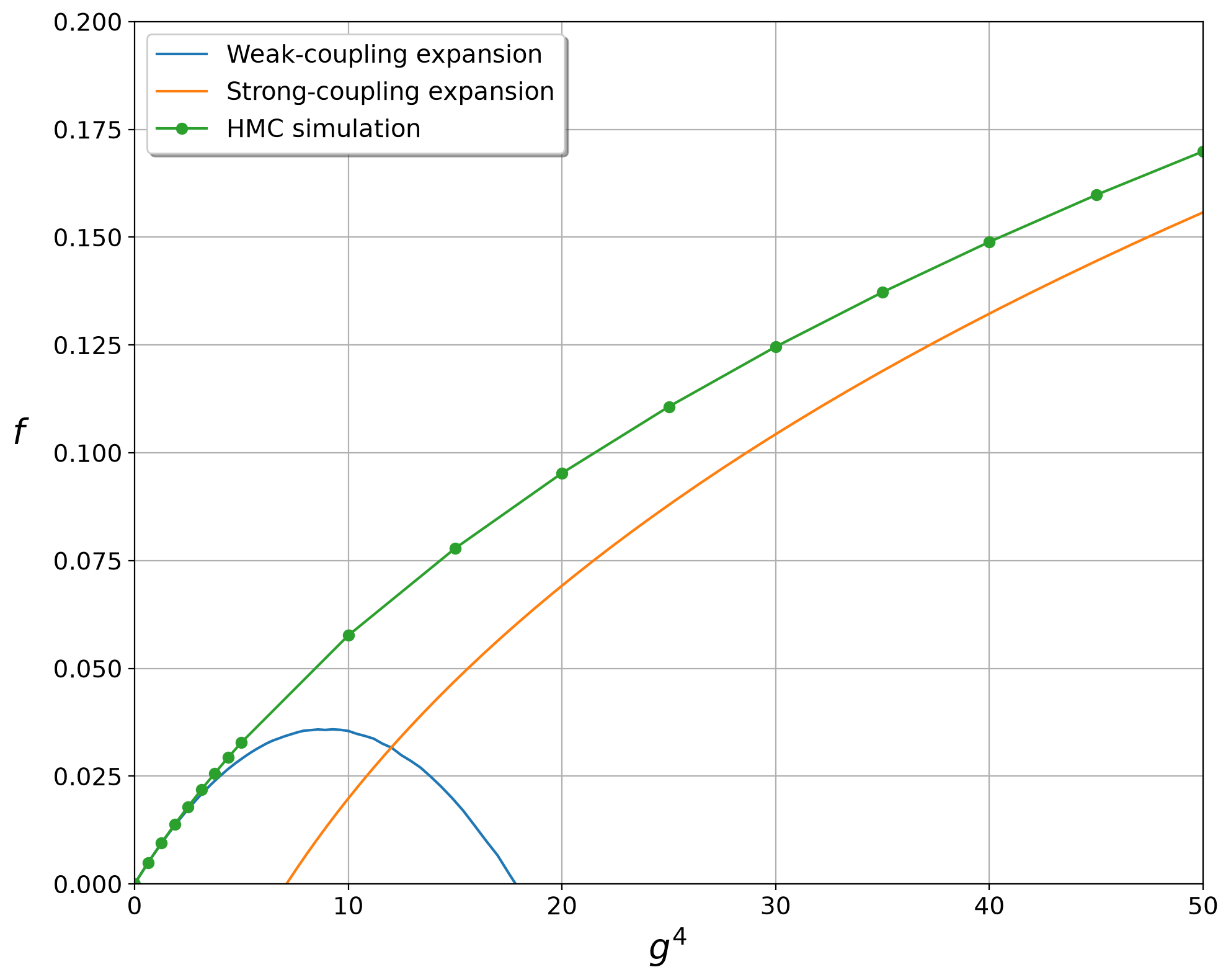}
    
    \includegraphics[width=8cm,height=5.5cm]{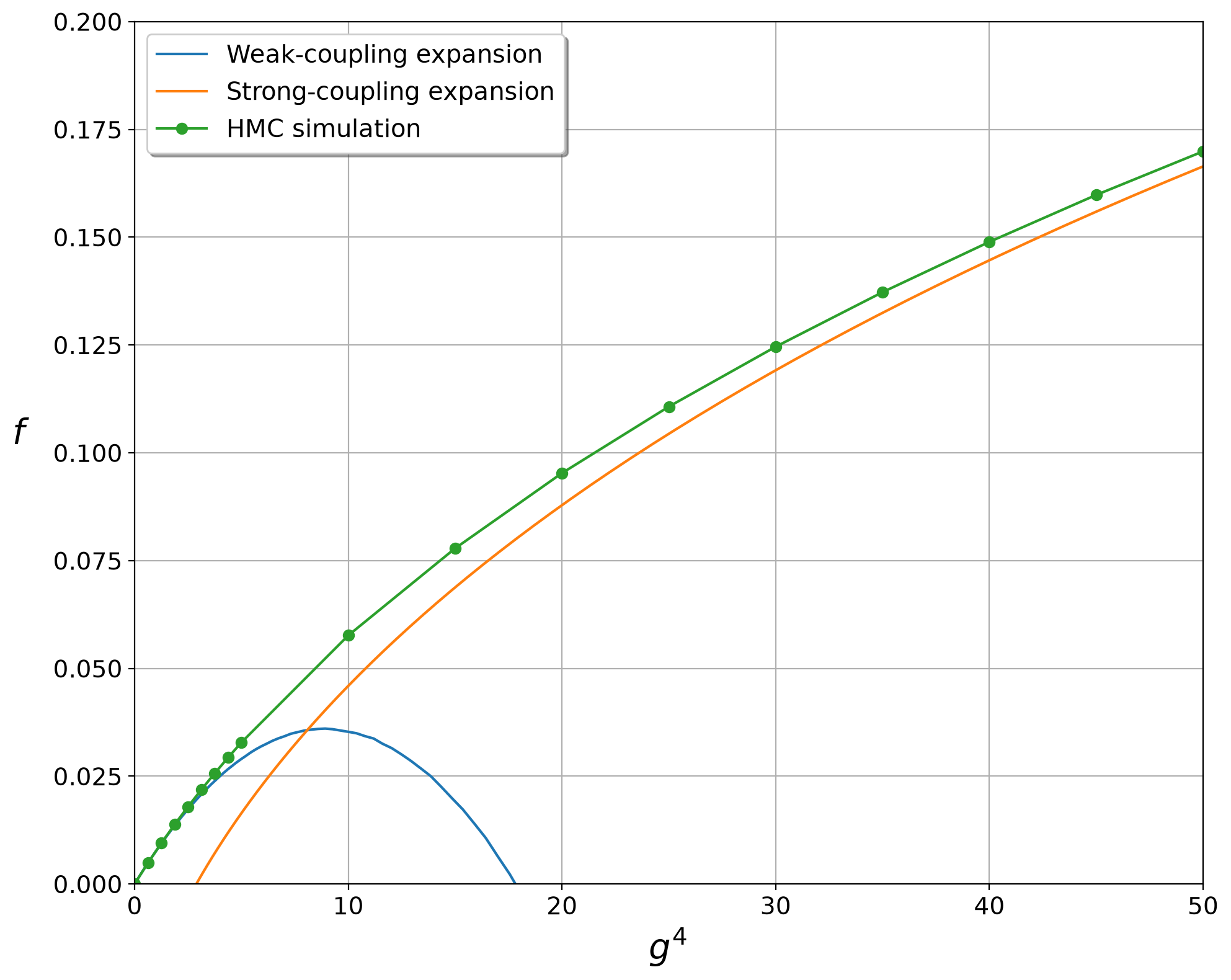}
    \hfill
    \includegraphics[width=8cm,height=5.5cm]{fg_2.png}
    \caption{Plots of analytical and computational results for the free energy per site \eqref{free-energy-per-site-def} for different orders of approximations in $1/g^2$ in strong coupling expansions for $d=2$. The top left picture corresponds to zeroth order approximation (i.e $\tilde{f}_0$ only), the top right - to second order, the bottom left - to the third, and the bottom right - to the fourth (as on Figure~\ref{fig:f(g)-comparison}). From the picture one can see that the constructed strong coupling series successive truncations approach the numerical result over the parameter range shown.}
    \label{fig:f(g)-comparison-diff-orders}
\end{figure}

It is instructive to plot how the described strong coupling series approaches the numerical results (Figure~\ref{fig:f(g)-comparison-diff-orders}). The quantitative estimation of free energy per site strong coupling approximation up to order $g^{-10}$ is shown on Figure~\ref{fig:f(g)-relative-errors}.

Over most of the strong-coupling range, the discrepancy between the analytical approximation and HMC is substantially larger than the HMC statistical uncertainty, indicating that the observed deviation is dominated by the finite-order truncation rather than numerical noise.

As a finite-size check, we repeated the simulations for several representative coupling values at smaller lattice sizes ($M=8,16,24$) and found that the resulting changes in the observables were smaller than their statistical uncertainties. We therefore neglect finite-size corrections at the level of accuracy considered here.

More detailed information about error estimation can be found in the original repository \cite{LatticePhi4_2023}.

\section{Application of the duality to the Ising model}
\label{appendix:duality-for-Ising}

It is insightful to examine the implications of the presented duality when applied to the Ising model, as this provides a valuable cross-check and clarifies the scope of the transformation. In this Appendix, we will follow the approach of path integral representation of the Ising model described, for instance, in \cite{Mussardo:SQFT}. In this Appendix we will assume that Ising model interaction matrix $J > 0$ - is symmetric and positively definite. Let us note that by adding a diagonal shift $\lambda \delta_{x,x^\prime}$ any matrix can be done positive for $\lambda>0$ large enough.

So, let us consider the statistical lattice model with partition function
and Hamiltonian of the form:
\[
\mathcal{Z}=\sum_{\left\{ \sigma_{x}=\pm1\right\} }e^{-\beta H\left[\left\{ \sigma_{x}\right\} \right]},\qquad H\left[\left\{ \sigma_{x}\right\} \right]=-\frac{1}{2}\sum_{x,x^{\prime}}J_{x,x^{\prime}}\sigma_{x}\sigma_{x^{\prime}},
\]
where the variables $\sigma_{x}$ represent spins defined on the lattice sites. We will consider the ferromagnetic regime, though
in principle the same could be done also for the antiferromagnetic
case. In the following, we will use the notation of Section~\ref{sect:duality-derivation}.
Using Hubbard--Stratonovich transformation:

\[
\frac{1}{(2\pi)^{N/2}}\int\frac{\prod_{x}d\phi(x)}{\sqrt{\det(\beta\hat{J})}}\exp\left(-\frac{1}{2}\sum_{x,x^{\prime}}(\beta J)_{x,x^{\prime}}^{-1}\phi(x)\phi(x^{\prime})+\sum_{x}\phi(x)\sigma_{x}\right)=\exp\left(\frac{\beta}{2}\sum_{x,x^{\prime}}J_{x,x^{\prime}}\sigma_{x}\sigma_{x^{\prime}}\right),
\]
we rewrite the Ising model as a field theory \cite{Mussardo:SQFT}:
\[
\mathcal{Z}=\frac{1}{(2\pi)^{N/2}}\int\frac{\prod_{x}d\phi(x)}{\sqrt{\det(\beta\hat{J})}}\exp\left(-\frac{1}{2}\sum_{x,x^{\prime}}(\beta J)_{x,x^{\prime}}^{-1}\phi(x)\phi(x^{\prime})+\sum_{x}\ln\left(2\cosh\phi(x)\right)\right).
\]

So, the action has a form:
\[
S[\phi]=\frac{1}{2}\sum_{x,x^{\prime}}(\beta J)_{x,x^{\prime}}^{-1}\phi(x)\phi(x^{\prime})-\sum_{x}\ln\left(2\cosh\phi(x)\right).
\]
Direct application of the duality (section~\ref{sect:duality-derivation}) to the action of this form is hindered, since the exponent of the interaction $-\sum_{x}\ln\left(2\cosh\phi(x)\right)$
lacks a standard Fourier transform even in the sense of distributions

\[
S[\phi]=\left(\frac{1}{2}\sum_{x,x^{\prime}}(\beta J)_{x,x^{\prime}}^{-1}\phi(x)\phi(x^{\prime})-\frac{\epsilon}{2}\sum_{x}\phi(x)^{2}\right)+\left(\frac{\epsilon}{2}\sum_{x}\phi(x)^{2}-\sum_{x}\ln\left(2\cosh\phi(x)\right)\right),
\]
for $\text{Re}\:\epsilon>0$ and small enough. Now we are able
to calculate the Fourier Transform of the exponential of each bracket
in the formula above. Applying the Plancherel Theorem, we obtain:
\begin{equation}
\begin{split}
        \mathcal{Z} =\frac{e^{\frac{N}{2\epsilon}}\epsilon^{-N/2}}{(2\pi)^{N/2}}\int\frac{\prod_{x}d\psi(x)}{\sqrt{\det(1-\epsilon\beta\hat{J})}} &\exp \left(-\frac{1}{2}\sum_{x,x^{\prime}}\left((\beta J)^{-1}-\epsilon\right)_{x,x^{\prime}}^{-1}\psi(x)\psi(x^{\prime})-\right. \\ & \left. \frac{1}{2\epsilon}\sum_{x}\psi(x)^{2}+\sum_{x}\ln\left(2\cos\frac{\psi(x)}{\epsilon}\right)\right),
\end{split}
\end{equation}

or, simplifying:
\begin{equation}
\mathcal{Z}=\frac{e^{\frac{N}{2\epsilon}}\epsilon^{-N/2}}{(2\pi)^{N/2}}\int\frac{\prod_{x}d\psi(x)}{\sqrt{\det(1-\epsilon\beta\hat{J})}}\exp\left(-\frac{1}{2\epsilon}\sum_{x,x^{\prime}}\left(1-\epsilon\beta J\right)_{x,x^{\prime}}^{-1}\psi(x)\psi(x^{\prime})+\sum_{x}\ln\left(2\cos\frac{\psi(x)}{\epsilon}\right)\right).
\end{equation}
This is exactly the form of dual action for the Ising model. Nevertheless,
a closer examination of the derived expression reveals that it is,
once again, the Ising field theory. More precisely, if one rewrites $\cos$ as the sum of exponents:
\[
\exp\left[\sum_{x}\ln\left(2\cos\frac{\psi(x)}{\epsilon}\right)\right]=\sum_{\left\{ \sigma_{x}=\pm1\right\} }\exp\left(\frac{i}{\epsilon}\sum_{x}\sigma_{x}\psi(x)\right),
\]
we obtain immediately:
\[
\mathcal{Z}=\frac{e^{\frac{N}{2\epsilon}}\epsilon^{-N/2}}{(2\pi)^{N/2}}\cdot(2\pi)^{N/2}\epsilon^{N/2}\sum_{\left\{ \sigma_{x}=\pm1\right\} }\exp\left(-\frac{1}{2}\sum_{x,x^{\prime}}\left(1-\epsilon\beta J\right)_{x,x^{\prime}}\sigma_{x}\sigma_{x^{\prime}}\right)=\sum_{\left\{ \sigma_{x}=\pm1\right\} } \exp\left(\frac{\beta}{2}\sum_{x,x^{\prime}}J_{x,x^{\prime}}\sigma_{x}\sigma_{x^{\prime}}\right),
\]
which is exactly the partition function of the original model. Firstly, we
see that the provided duality leads us to a correct, though trivial, result for the case of the Ising model. Secondly, its application is equivalent
to the addition of a trivial term to the initial Hamiltonian. Thus, the duality transformation maps the Ising model onto itself, demonstrating the consistency of the formalism. Whether this triviality is a general property of all theories with discrete degrees of freedom (such as the Potts model) or a specific feature of the Ising model remains an open question for future study. For many other theories, such as lattice $\phi^{4}$, it yields a non-trivial
transformation, as we have seen in Section~\ref{sect:duality-derivation}.


\end{appendices}

\end{document}